\documentclass[]{spie}  

\def\cm{{\rm\thinspace cm}}

\def\erg{{\rm\thinspace erg}}
\def\eV{{\rm\thinspace eV}}

\def\km{{\rm\thinspace km}}

\def\m{{\rm\thinspace m}}

\def\Msun{\hbox{$\rm\thinspace M_{\odot}$}}

\def\s{{\rm\thinspace s}}

\def\cmsq{\hbox{$\cm^2\,$}}

\def\ergpcmsqps{\hbox{$\erg\cm^{-2}\s^{-1}\,$}}

\def\h18{\hbox{H1821$+$643\,}}

\usepackage{amsmath,amsfonts,amssymb}
\usepackage{graphicx}
\usepackage[colorlinks=true, allcolors=blue]{hyperref}

\usepackage[english]{babel} 
\usepackage{blindtext}

\title{Overview of the Advanced X-ray Imaging Satellite (AXIS)}

\author[a,b]{Christopher S. Reynolds}
\author[c]{Erin A. Kara}
\author[a,b]{Richard F. Mushotzky}
\author[d]{Andrew Ptak}
\author[e]{Michael J. Koss}
\author[d]{Brian J. Williams}
\author[f,g,h]{Steven W. Allen}
\author[i,j,k,l]{Franz E. Bauer}
\author[c]{Marshall Bautz}
\author[m]{Arash Bodaghee} 
\author[c]{Kevin B. Burdge}
\author[n]{Nico Cappelluti} 
\author[d]{Brad Cenko}
\author[o]{George Chartas}
\author[d]{Kai-Wing Chan}
\author[p]{L\'ia Corrales} 
\author[q]{Tansu Daylan}
\author[r]{Abraham D. Falcone}
\author[s]{Adi Foord}
\author[c]{Catherine E. Grant}
\author[t,u]{M\'elanie Habouzit}
\author[v,w]{Daryl Haggard} 
\author[f]{Sven Herrmann}
\author[d]{Edmund Hodges-Kluck}
\author[x]{Oleg Kargaltsev}
\author[p]{George W. King}
\author[y]{Marina Kounkel}
\author[z]{Laura A. Lopez}
\author[aa,bb,cc]{Stefano Marchesi}
\author[c]{Michael McDonald}
\author[s]{Eileen Meyer}
\author[c]{Eric D. Miller}
\author[c]{Melania Nynka}
\author[d]{Takashi Okajima}
\author[dd]{Fabio Pacucci}
\author[ee]{Helen R. Russell}
\author[ff]{Samar Safi-Harb}
\author[gg]{Keivan G.\ Stassun}
\author[dd]{Anna Trindade Falc\~ao}
\author[hh]{Stephen A. Walker}
\author[jj]{Joern Wilms}
\author[d,kk]{Mihoko Yukita}
\author[d]{William W. Zhang}
\author[.]{on behalf of the AXIS Science Community}

\affil[a]{\footnotesize Department of Astronomy, University of Maryland, College Park, MD20742, USA}
\affil[b]{Joint Space Science Institute (JSI), University of Maryland, College Park, MD20742, USA}
\affil[c]{MIT Kavli Institute for Astrophysics and Space Research and Department of Physics, MIT, Cambridge, MA 02139, USA}
\affil[d]{NASA Goddard Space Flight Center, Greenbelt, MD20771, USA}
\affil[e]{Eureka Scientific, 2452 Delmer Street Suite 100, Oakland, CA 94602-3017, USA}
\affil[f]{Kavli Institute for Particle Astrophysics and Cosmology, Stanford University, CA 94305, USA}
\affil[g]{SLAC National Accelerator Laboratory, 2575 Sand Hill Road, Menlo Park, CA 94025, USA}
\affil[h]{Dept. of Physics, Stanford University, 382 Via Pueblo Mall, CA 94305, USA}
\affil[i]{Instituto de Astrof{\'{\i}}sica, Facultad de F{\'{i}}sica, Pontificia Universidad Cat{\'{o}}lica de Chile, Campus San Joaquín, Av. Vicuña Mackenna 4860, Macul Santiago, Chile, 7820436}
\affil[j]{Centro de Astroingenier{\'{\i}}a, Facultad de F{\'{i}}sica, Pontificia Universidad Cat{\'{o}}lica de Chile, Campus San Joaquín, Av. Vicuña Mackenna 4860, Macul Santiago, Chile, 7820436}
\affil[k]{Millennium Institute of Astrophysics, Nuncio Monse{\~{n}}or S{\'{o}}tero Sanz 100, Of 104, Providencia, Santiago, Chile}
\affil[l]{Space Science Institute, 4750 Walnut Street, Suite 205, Boulder, Colorado 80301}
\affil[m]{Dept. of Chemistry, Physics and Astronomy, Georgia College \& State University, Milledgeville, GA 31061, USA}
\affil[n]{Department of Physics, University of Miami, Coral Gables, FL 33124, USA}
\affil[o]{Department of Physics and Astronomy, College of Charleston, Charleston, SC, 29424, USA}
\affil[p]{Department of Astronomy, University of Michigan, Ann Arbor, MI 48109, USA}
\affil[q]{Department of Physics and McDonnell Center for the Space Sciences, Washington University, One Brookings Drive, St. Louis, MO 63130-4899, USA}
\affil[r]{Dept. of Astronomy \& Astrophysics, Pennsylvania State University, University Park, PA 16803, }
\affil[s]{Dept. of Physics, University of Maryland Baltimore County, Baltimore, MD 21250, USA}
\affil[t]{Zentrum für Astronomie der Universit\"at Heidelberg, ITA, Albert-Ueberle-Str. 2, D-69120 Heidelberg, Germany}
\affil[u]{Max-Planck-Institut f\"ur Astronomie, MPIA, K\"onigstuhl 17, D-69117 Heidelberg, Germany}
\affil[v]{Dept. of Physics, McGill University, 3600 rue University, Montr\'{e}al, QC H3A2T8, Canada}
\affil[w]{Trottier Space Institute at McGill, 3550 rue University, Montr\'{e}al, QC H3A2A7, Canada}
\affil[y]{Dept. of Physics and Astronomy, University of North Florida, 1 UNF Dr, Jacksonville, FL, 32224}
\affil[x]{Department of Physics, The George Washington University, 725 21st Street NW, Washington, DC 20052, USA}
\affil[z]{Dept. of Astronomy, The Ohio State University, Columbus, Ohio 43210, USA}
\affil[aa]{Dipartimento di Fisica e Astronomia, Università di Bologna, via Gobetti 93/2, I-40129 Bologna, Italy}
\affil[bb]{Department of Physics and Astronomy, Clemson University,  Kinard Lab of Physics, Clemson, SC 29634, USA}
\affil[cc]{INAF - Osservatorio di Astrofisica e Scienza dello Spazio di Bologna, Via Piero Gobetti, 93/3, 40129, Bologna, Italy}
\affil[dd]{Center for Astrophysics $\vert$ Harvard \& Smithsonian, Cambridge, MA 02138, USA}
\affil[ee]{School of Physics \& Astronomy, University of Nottingham, University Park, Nottingham NG7 2RD, UK}
\affil[ff]{Dept. of Physics \& Astronomy, The University of Manitoba, Winnipeg, MB R3T2N2, Canada}
\affil[gg]{Department of Physics and Astronomy, Vanderbilt University, Nashville, TN 37235, USA}
\affil[hh]{Department of Physics and Astronomy, The University of Alabama in Huntsville, Huntsville, AL 35899, USA}
\affil[ii]{Center for Space Science and Technology, University of Maryland, Baltimore County, Baltimore, MD 21250, USA}
\affil[jj]{Dr. Karl Remeis-Observatory and Erlangen Centre for Astroparticle Physics, Friedrich-Alexander Universität Erlangen-Nürnberg, Sternwartstr. 7, 96049 Bamberg, Germany}
\affil[kk]{The William H. Miller III Dept. of Physics and Astronomy, The Johns Hopkins University, Baltimore, MD 21218, USA}

\authorinfo{Further author information: (Send correspondence to C.S.R.)\\C.S.R.: E-mail: creynold@umd.edu, Telephone: 1 301 405 3001}

\begin{document} 
\maketitle

\begin{abstract}
The Advanced X-ray Imaging Satellite (AXIS) is a Probe-class concept that will build on the legacy of the Chandra X-ray Observatory by providing low-background, arcsecond-resolution imaging in the 0.3-10 keV band across a 450 arcminute$^2$ field of view, with an order of magnitude improvement in sensitivity.  AXIS utilizes breakthroughs in the construction of lightweight segmented X-ray optics using single-crystal silicon, and developments in the fabrication of large-format, small-pixel, high readout rate CCD detectors with good spectral resolution, allowing a robust and cost-effective design.  Further, AXIS will be responsive to target-of-opportunity alerts and, with onboard transient detection, will be a powerful facility for studying the time-varying X-ray universe, following on from the legacy of the Neil Gehrels (Swift) X-ray observatory that revolutionized studies of the transient X-ray Universe.  In this paper, we present an overview of AXIS, highlighting the prime science objectives driving the AXIS concept and how the observatory design will achieve these objectives.
\end{abstract}


\section{INTRODUCTION}

From the discovery of the Galilean moons \cite{galileo:1610a} to the identification of galaxies at $z>13$ ($<400$ million years after the Big Bang) \cite{robertson:22a}, major advances in astronomy go hand in hand with the development of high-spatial resolution, high-throughput telescopes.  Superior spatial resolution translates into superior sensitivity, allowing us to probe deeper, fainter, and further.   This is as true in the X-ray band as it is in the optical/infrared frequencies that have traditionally led astronomy.

Today's premiere high spatial resolution X-ray observatory, the {\it Chandra X-ray Observatory} (CXO), has made seminal contributions across all areas of astronomy and astrophysics \cite{wilkes:22a}.  The CXO deep surveys\cite{luo:17a} transformed our understanding of supermassive black hole (SMBH) evolution across cosmic time, charting the growth of SMBHs from a billion years after the Big Bang to the present day and, in concert with the {\it Hubble Space Telescope} and ground-based facilities, the connection of SMBH growth with star formation\cite{lehmer:08a}.   CXO observations of the hot, diffuse intracluster medium (ICM) of galaxy clusters provided the first direct evidence that SMBHs were strongly influencing their super-Galactic environment and accelerated our understanding of active galactic nucleus (AGN) feedback\cite{fabian:00a}.   CXO studies of star-forming regions such as the Orion Complex\cite{preibisch:05a} profoundly influenced our view of the processes occurring in young protoplanetary systems, highlighting that young stars can be very X-ray active and that this activity can be important for the accretion-driven growth of the star as well as the properties of the planets that may be forming.  These are but a few examples of CXO's profound legacy. 

CXO was designed in the early 1980s and deployed nearly a quarter century ago. Today, there is a growing gap between our capabilities in the X-ray band and those in the optical and near/mid-IR bands.  This mismatch impacts our ability to address the most pressing questions in contemporary astronomy and astrophysics.  JWST is discovering candidate AGNs at $z>10$ (less than half a billion years after the Big Bang) but a rigorous confirmation that these are truly accreting $10^6-10^7\Msun$ SMBHs, and characterization of their properties requires X-ray observations at a depth currently not possible\cite{bogdan:23a,maiolino:23a,goulding:23a}. In the low-redshift Universe, CXO and XMM-Newton have firmly established the action of jet-mediated AGN feedback on galaxy cluster scales, but it is largely unknown whether scaled-down versions of these processes operate in the interstellar and circumgalactic medium of individual galaxies. Even closer to home, surprising findings on the atmospheres of exoplanets, for example the discovery by JWST that the planets TRAPPIST-1b and TRAPPIST-1c are likely ``bare rocks'', highlights the importance of exoplanet atmospheric escape driven by the coronal extreme-UV and X-ray activity of the host star, again mostly beyond the reach of current X-ray observatories.

The {\it Advanced X-ray Imaging Satellite} (AXIS), a response to NASA's Astrophysics Probe Explorer (APEX) program, will be a critical part of the 2030s astrophysics landscape.  With $1-2$\,arcsecond quality imaging across a 24\,arcminute diameter field-of-view and a sensitivity $>10\times$ that of CXO, AXIS offers X-ray capabilities that are well-matched in depth to JWST and the near-future ground- and space-based facilities that will dominate the next 15 years of astronomical research.  Furthermore, AXIS will be a powerful time-domain observatory, with $\sim 80\times$ the sensitivity of {\it Swift}\cite{gehrels:04a}, acting as the primary discovery engine for supernova shock breakouts and other fast X-ray transients, and responding in $<2$\,hours to any community alerts deemed sufficiently interesting. AXIS directly addresses fundamental questions that span all three themes of the Astro2020 Decadal Survey report, {\it Pathways to Discovery in Astronomy and Astrophysics for the 2020s}.  AXIS has been enabled by developments in segmented X-ray optics technology using single-crystal silicon mirrors; these developments in turn build on the colossal investments made by the semiconductor industry in the affordable production and manipulation of single-crystal silicon. {\bf With these specific technological advances, and by adopting the class-C risk posture defined by the APEX program, it is now possible to build an observatory that is substantially more powerful than CXO at a fraction of the cost.}

AXIS will be a major community facility for the 2030s, with $>70\%$ of the observing time driven by General Observers worldwide. Through synergies with other facilities in the electromagnetic (e.g., ALMA, ELTs, Roman, Rubin, and ngVLA), gravitational wave media (e.g., Einstein, LISA, IPTA30), and neutrino realm (e.g. ICECUBE-2), AXIS will be a capable facility for addressing time-domain and multi-messenger (TDAMM) astrophysics, a major Astro2020 theme and NASA priority area.  In this paper, we present an observatory-level overview of AXIS.  We highlight how the science drives the primary instrument and observatory requirements before providing top-level descriptions of the observatory architecture and the mission implementation. 

\section{AXIS IN A NUTSHELL}

The scientific questions that motivate AXIS (Section~\ref{sec:science}) demand exquisite sensitivity to both point- and diffuse-sources in the 0.3-10\,keV band, as well as rapid response to transient phenomena.  Every aspect of the AXIS design is geared towards achieving these capabilities.

AXIS is a single-instrument observatory.  The X-ray Mirror Assembly (XMA) focuses X-rays onto an array of charge-coupled devices (CCDs) within the Focal Plane Array (FPA), resulting in position-, energy-, and time-tagged photon detections that can be binned into X-ray images, spectra, and light-curves.  The principal characteristics of the observatory are presented in ~\ref{tab:axis_params}.  

\begin{table}[t]
\caption{Characteristics of the Baseline AXIS Mission}\label{tab:axis_params}
\label{tab:baseline}
\begin{center}       
\begin{tabular}{|l|l|} 
\hline\hline
Bandpass & 0.3--10\,keV \\\hline
Spatial resolution 1keV (HPD) & 1.5$''$ (on-axis) \\
& 1.6$''$ (FoV-average) \\\hline 
Effective area at 1keV & $4200\cmsq$ (on-axis) \\
& $3600\cmsq$ (FoV-average) \\\hline
Effective area at 6keV & $830\cmsq$ (on-axis) \\
& $570\cmsq$ (FoV-average) \\\hline
Field-of-View & $24$\,arcmin diameter \\\hline
Detector readout cadence & $>5$\,frames/s \\\hline
Energy resolution & 70\eV (at 1keV) \\
& 150\eV (at 6keV)\\\hline
Response time to community alerts & on-source in $<2$\,hours \\
Response to detected transients & to community in $<10$\,mins \\\hline
Orbit &Circular Low-Earth Orbit \\
& $610-680\km$ altitude \\
& $<8$\,degree inclination \\\hline
Mission lifetime & 5\,yr prime mission\\
\hline\hline
\end{tabular}
\end{center}
\end{table} 

\begin{figure} [t]
\begin{center}
\begin{tabular}{c} 
\includegraphics[width=0.8\textwidth]{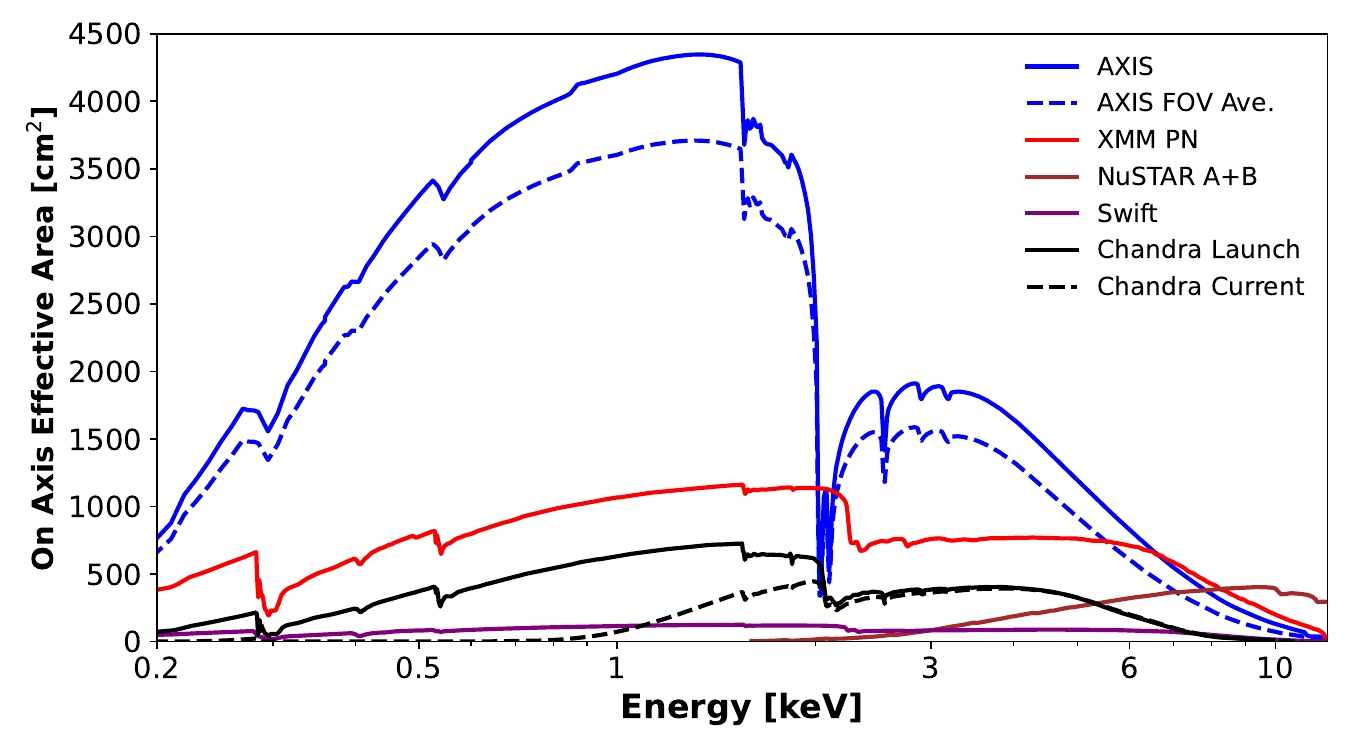}
\vspace{-0.3cm}
\end{tabular}
\end{center}
\caption{Mirror effective area for AXIS (blue) compared to XMM-PN (red), NuSTAR (A and B telescopes combined; brown),  CXO at launch (black) and CXO in 2023 and hence affected by contamination buildup (dashed).}\label{fig:effarea}
\end{figure}

Sensitivity is key to achieving the scientific goals of AXIS, and this demands large collecting power and high spatial resolution imaging to avoid source confusion and disambiguate point sources from background emission.  The XMA produces an on-axis spatial resolution of $<1.5''$\ (half-power diameter; HPD), significantly better than all previous X-ray observatories except CXO. AXIS has a 24$'$ diameter field-of-view (FoV) and, unlike CXO, experiences only a weak degradation of spatial resolution away from the optical axis, enabling a field-of-view average resolution of $<1.6''$\ (HPD), and an order-of-magnitude more effective resolution elements in the FoV as compared with CXO.  As shown in Figure~\ref{fig:effarea}, the on-axis effective area of AXIS (accounting for all losses due to finite reflectivity, detector quantum efficiencies, and absorption by the contamination blocking filter) is $3300\cmsq$ at 0.5\,keV ($9\times$ CXO), $4200\cmsq$ at 1\,keV ($6.6\times$ CXO), and $830\cmsq$ at 6\,keV ($4.3\times$ CXO).  These comparisons are made against the CXO Advanced CCD Imaging Spectrometer (ACIS) in its ``pristine state'' just after deployment in July-1999; the on-orbit build-up of molecular contamination on the ACIS optical blocking filters has very strongly impacted the $<1$\,keV response of CXO/ACIS and thus the comparison of AXIS with today's CXO/ACIS is far more dramatic at low-energies.

The AXIS detector, the FPA, consists of a $2\times 2$ array of back-illuminated MIT/Lincoln-Lab X-ray CCDs cooled to an operating temperature of $-90^\circ{\rm C}$ and read out with a cadence of $<0.2$\,s and low noise of $<3.0e^-$. With each CCD possessing $1440\times 1440$ $24\mu$m-square pixels, the AXIS focal plane array covers the entire 24\,arcmin FoV with 0.5\,arcsec-square pixels that over-sample the spatial resolution by a factor of 2--3, thereby permitting high-quality centroiding of detected sources. Learning the lessons from CXO, significant measures have been taken to prevent the build-up of molecular contamination that can degrade the low-energy X-ray sensitivity, both during construction/testing and on-orbit.  In addition to careful control of potential contamination sources throughout the whole system, the CCD will be housed in a vacuum enclosure protected from the rest of the system by a warm ($+20^\circ{\rm C}$) contamination blocking filter.  This will permit AXIS to maintain sensitivity to low-energy X-rays (0.3-1.0\,keV) throughout the five-year prime mission.  

\begin{figure}[t]
\begin{center}
\begin{tabular}{c} 
\hspace{-2cm}
\includegraphics[height=7cm]{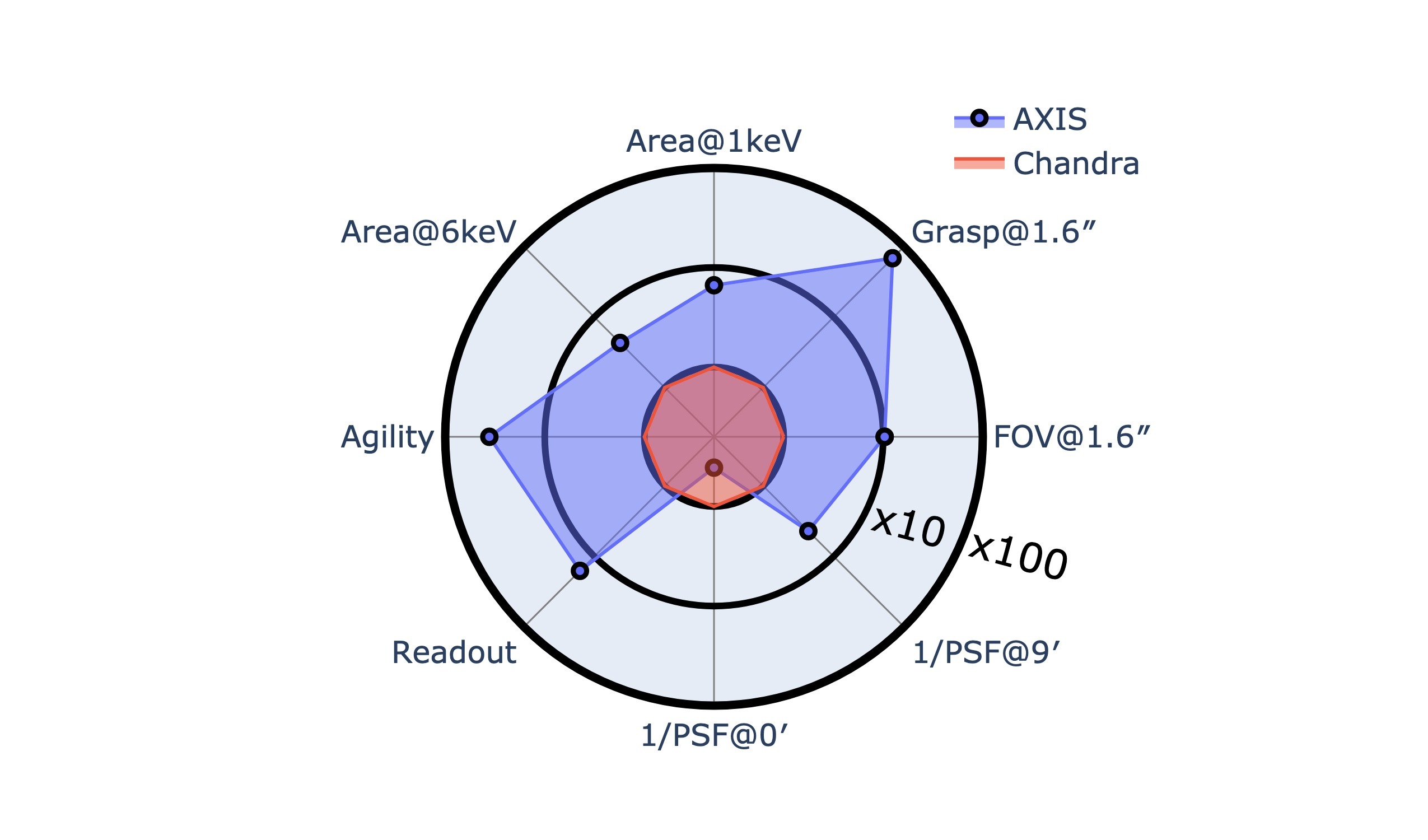}
\hspace{-1cm}
\includegraphics[height=6.5cm]{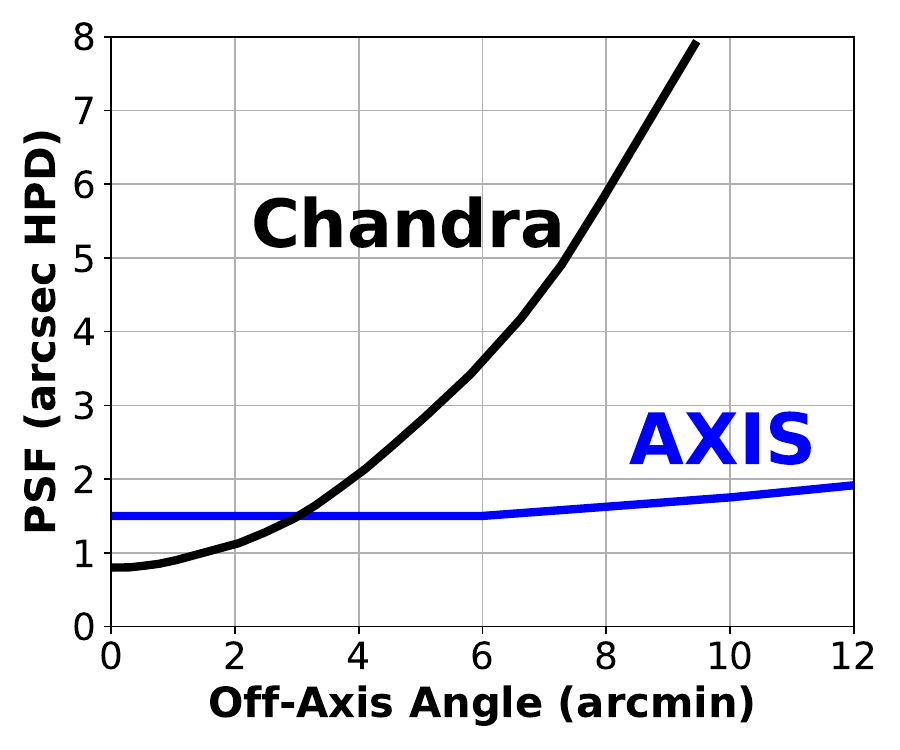}
\end{tabular}
\end{center}
\vspace{-0.2cm}
\caption{{\em Left}: AXIS radial logarithmic plot which shows how AXIS (blue) compares with the performance of CXO in several different capabilities critical for science objectives (red 1 on all axes); concentric circles denote $\times$1, $\times$10, and $\times$100 relative improvements (higher/lower depending of the parameter).  With a combination of its high angular resolution across a large field of view, large increase in effective area, agility to respond quickly to transient events, and high throughput, AXIS opens an order of magnitude or more discovery space for its science objectives with only a small decrease in on-axis spatial resolution. {\em Right}: AXIS and CXO PSF with off-axis angle.  The optical design of the XMA permits high-spatial resolution across the entire field of view. }\label{fig:radar}
\end{figure}

A broader comparison between AXIS and CXO is made in Figure~\ref{fig:radar}.  Crucially, the combination of effective area and the large field of view that is imaged with high-resolution gives AXIS almost two orders of magnitude greater high-resolution survey grasp (i.e., field of view $\times$ effective area) than CXO.  

AXIS will be placed in a low-inclination ($<8$\,degree) low-Earth orbit (LEO).  The primary drivers for this choice are the need to maintain rapid, on-demand (low data-rate) communications for time-domain science, and to obtain a low radiation environment which gives a low detector background and slows detector degradation.  Despite being in LEO, AXIS maintains an observing efficiency of $>74\%$\footnote{During normal operation, i.e., not including calibration time and safeholds} through its ability to rapidly slew between targets.

\section{THE SCIENCE OF AXIS}\label{sec:science}

As evidenced by the nearly 10,000 refereed journal articles written over the last 24 years using CXO observations, the discovery space for high spatial resolution X-ray astronomy is vast. AXIS will dramatically expand this discovery space by going wider, deeper, and further than ever before. The prime mission AXIS Science case is organized around four science pillars:

\subsection{COSMIC ECOSYSTEMS: The Formation and Growth of Supermassive Black Holes}

{\it What are the seeds of supermassive black holes (SMBHs)? What is the role of mergers in SMBH growth?} 

SMBHs that range in mass from 10$^{6}$ - 10$^{9}$ M$_\odot$ are ubiquitous in massive galaxies, and yet their origin and early growth remain a mystery. More staggering than their sheer existence is the fact that some of the largest of these black holes existed in the early universe, within the first billion years after the Big Bang. 

However, the origins of SMBHs are still unknown.  The problem has been studied extensively by theorists using both semi-analytic galaxy formation and cosmological hydrodynamic simulations. The two most popular scenarios to explain the rapid assembly of the quasar population by 1\,Gyr, which bracket the extremes of seeding models, are (1) the “light” seed scenario, in which stellar remnants of Population III stars ($>10^{2-3}$ M$_\odot$, \cite{mcconnell:13}) grow extremely rapidly (at super-Eddington rates), or (2) the “heavy” seed scenario in which massive black holes are created in the direct collapse of massive metal-free clouds in protogalactic cores ($>10^{4}$ M$_\odot$, \cite{ferrarese:00, gultekin:09, magorrian:98, tremaine:02}). Because of the ambiguity between black hole seeds and their accretion histories, both scenarios can explain the population statistics of bright-massive quasars up to z$\sim 7$, the horizon for current X-ray observations. 

\begin{figure} [t]
\begin{center}
\begin{tabular}{c} 
\includegraphics[width=\textwidth]{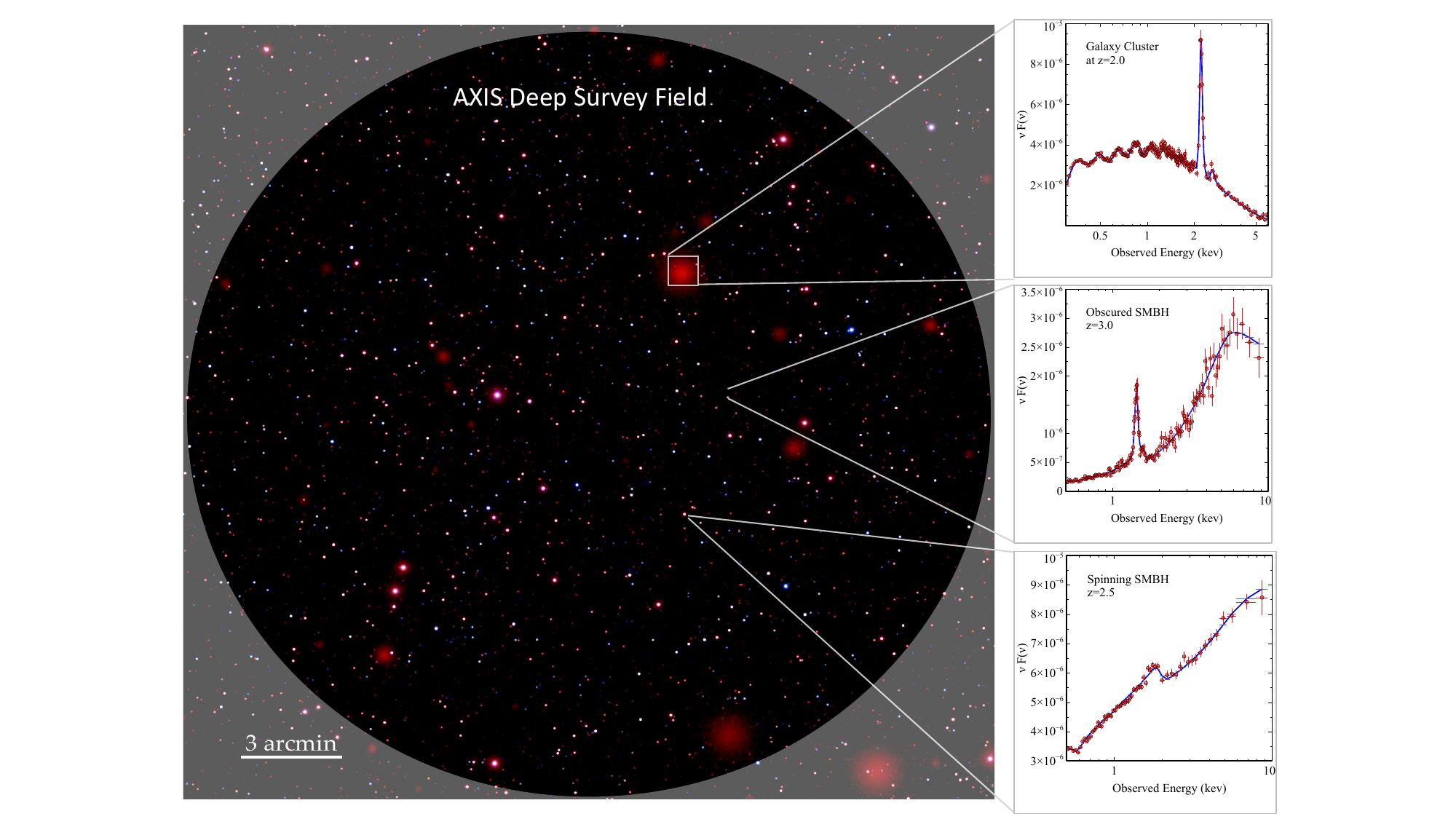}\\
\includegraphics[width=0.8\textwidth]
{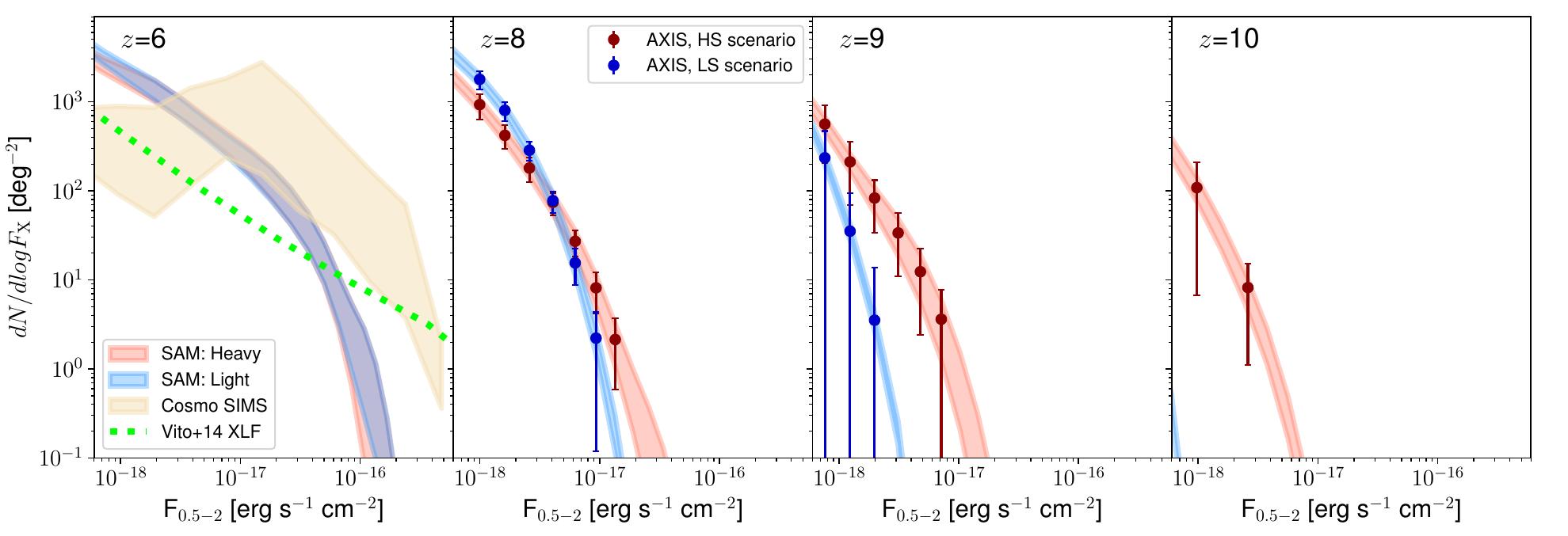}
\end{tabular}
\end{center}
\caption{
\textit{Top}: Tricolor (red: 0.5--2\,keV; green: 2--4.5\,keV blue: 4.5--7\,keV) image of the simulated AXIS 7.0 Ms Deep Survey Field. The image has been lightly smoothed for display purposes to increase the visibility of the point sources.  The AXIS surveys will enable a range of investigations with unprecedented statistics: an (incomplete) list of examples is highlighted in the three panels on the right. In the top one, we present the spectrum of a $z$=2 cluster of galaxies; in the central one, we report the spectrum of a $z$=3 SMBH accreting in a heavily obscured phase; and in the bottom panel we report the spectrum of a spinning SMBH at $z$=2.5. 
\textit{Bottom}: Differential number counts of high-z AGN and their evolution as a function of z, L$_{X}$, F$_{X}$. The gold band is the dispersion of results from numerical simulations\cite{habouzit22a}, the green dotted line shows the extrapolation at z$\sim 6$ of the X-ray luminosity function measured at z$\sim 4$ using the deepest \textit{Chandra} fields currently available\cite{vito14}, while the light red and blue curves are predictions for heavy and light seed models, respectively\cite{ricarte18}.}\label{fig:deepxlf}
\end{figure}

The AGN population statistics predicted by light- and heavy-seed scenarios begin to strongly diverge beyond a redshift of $z=8$ ($<640$ million years after the Big Bang). In essence, if the main channel for black hole growth is heavy seeds, these sources will become AGN earlier in the universe (i.e. at $z=10$), whereas light seeds will be fainter for longer, and will only approach AGN luminosities at $z=8$. The AXIS Deep and Wide extragalactic surveys will be conducted in fields with good JWST and Roman coverage, and will conduct an X-ray census of AGN with unprecedented depths; limiting fluxes will be $F_{0.5-2\,{\rm keV}}<3\times 10^{-17}\ergpcmsqps$ across 7000\,arcmin$^2$, and $F_{0.5-2\,{\rm keV}}<4\times 10^{-18}\ergpcmsqps$ across 400\,arcmin$^2$.   This will allow a measurement of the X-ray Luminosity Function of AGN up to $z=10$ with sufficient sensitivity and survey volume to distinguish the SMBH seeding scenario (Figure~\ref{fig:deepxlf}).   

As these SMBHs evolve over cosmic time, they grow from the accretion of gas, which can be significantly enhanced when galaxies collide and merge, and direct mergers of SMBHs. However, the relative importance of accretion and merger-driven growth remains largely unknown. The high spatial resolution imaging of the AXIS Deep and Wide Extragalactic Surveys will also survey dual accreting black holes creating an unbiased census of interacting pairs in order to uniquely measure the enhancement of black hole accretion through mergers. 

AXIS will determine the importance of galactic mergers in the growth of massive black holes by precisely measuring the dual AGN fraction (AGN with a companion separated by $7-30$\,kpc) throughout cosmic time from z=0 to z=3.5 through the extragalactic surveys. Dual AGN, in the form of pairs of AGN in merging galaxies, represent periods when both SMBHs are active, and are unique observational flags of merger-driven supermassive black hole accretion and growth. To date, several studies have used the superior angular resolution of CXO to discover closely separated dual AGN \cite{koss:12,foord:19,foord:21}. However, we have reached the limits to what CXO can achieve: to date, fewer than 50 dual AGN have been detected and confirmed in X-rays \cite{chen:22}, and nearly all are restricted to the local universe (z$<0.1$) due to lack of sensitivity at $\sim 1$\,arcsec. 

The AXIS deep and wide extragalactic surveys will identify and characterize $>20,000$ AGN ($>5000$ of those having $z>2$). With a field-of-view average PSF of $<1.50$\,arcsec, AXIS will be able to find the dual AGN sub-population with separations $>7$\,kpc. Depending upon the importance of mergers for SMBH growth, we expect to find between 200--1000 dual AGN (50--250 at $z>2)$.  Mapping the dual fraction as a function of redshift will determine the role (or lack thereof) that galaxy mergers play in enhancing SMBH growth across cosmic time. 

\subsection{COSMIC ECOSYSTEMS: Feedback in Galaxies Across All Mass Scales}

{\it How does stellar feedback affect the evolution of low-mass galaxies? Is there evidence for AGN feedback in individual gas-rich galaxies? When in cosmic time did AGN feedback begin?} 

Galaxies form and evolve through a vigorous cosmic struggle.  Gravity drives the assembly of galaxies by drawing primordial gas into dark matter halos and collapsing cold clouds into star formation. Yet we observe that only up to 15--20\% of baryons are actually converted to stars\cite{weshsler:18a}.  This inefficiency across the galaxy mass scale must be the result of high-energy processes around stars and massive black holes.  Stellar feedback, via stellar winds and supernovae, heats and disperses cold gas clouds and, collectively, inflates large hot gas bubbles that lift hot gas and metals into the galactic halo \cite{veilleux:05,heckman:17}.  In massive galaxies, more powerful AGN feedback, in the form of radiation, winds, and relativistic jets, is required to drive large outflows, suppress cooling of hot atmospheres, and produce a correlative slowing of black hole activity \cite{fabian:12}.  Feedback indelibly shapes galaxy evolution across the galaxy mass function, back to the peak of galaxy growth at $z\sim 2$ and beyond.  

Although cosmological structure formation simulations rely on prescriptions for stellar and AGN-driven feedback to match observations, direct causal evidence and a mechanistic understanding of feedback remain inconclusive.  X-ray observations are key because they reveal both the cause of feedback originating from small scales (e.g. stellar winds, AGN jet-driven cavities, supernova remnants) and the corresponding effect on large scales (e.g. the heating and metal enrichment of the ISM across the galaxy halo). Arcsecond spatial resolution in X-rays is essential for disentangling the complex interactions of winds, radiation, jets and supernovae with the surrounding gas.  But, only when combined with high throughput can we reach the key scales of individual HII regions, star clusters, wind shocks, and jet-blown bubbles across the galaxy population and trace the establishment of feedback back to cosmic noon. 

AXIS will study how stellar winds and supernovae shape galaxies by measuring how efficiently their energy couples to the interstellar medium in thousands of massive star clusters spanning a range of masses, ages, metallicities, and environments. The injection of energy and momentum by stars fundamentally changes the properties of the galaxy at large (e.g., its shape and structure, and the rate and efficiency of star formation). Stellar winds and supernovae from massive star clusters inflate parsec-scale hot bubbles in the surrounding interstellar medium (ISM). Individual bubbles disrupt natal clouds, triggering star formation on their periphery, and dispersing gas and metals that become fuel for the next generation of stars. Collections of these bubbles build on each other and, in the extreme cases of starburst nuclei, form superheated gas that can drive galactic winds that eject much of the ISM gas into galactic halos.  

\begin{figure} [t]
\begin{center}
\begin{tabular}{c} 
\includegraphics[width=0.9\textwidth]{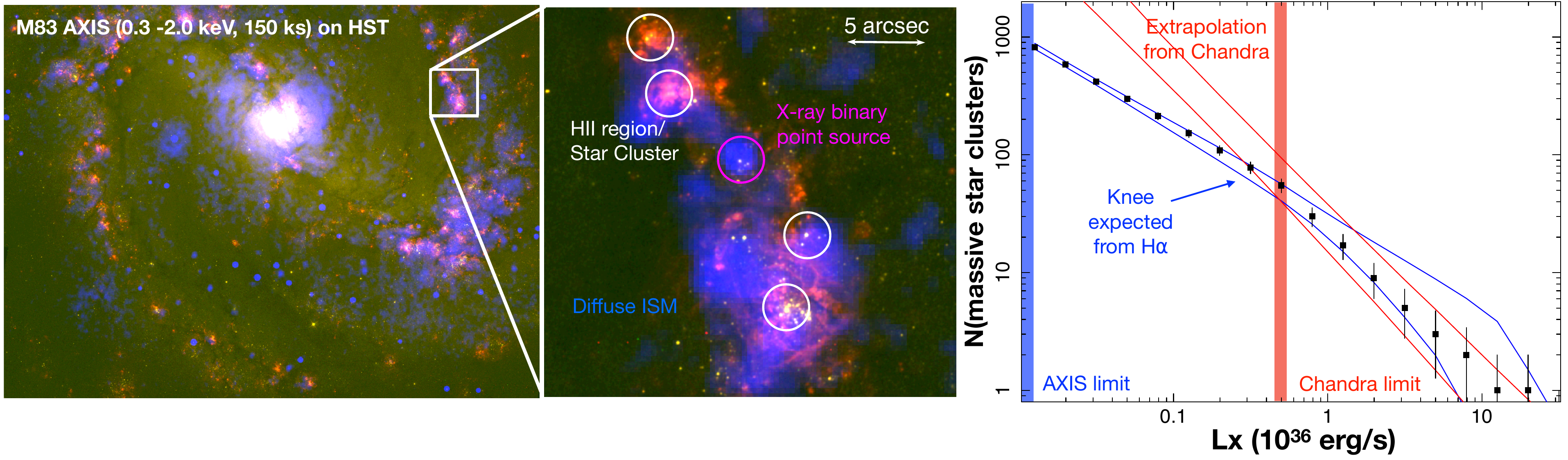}\\
\includegraphics[width=0.3\textwidth]{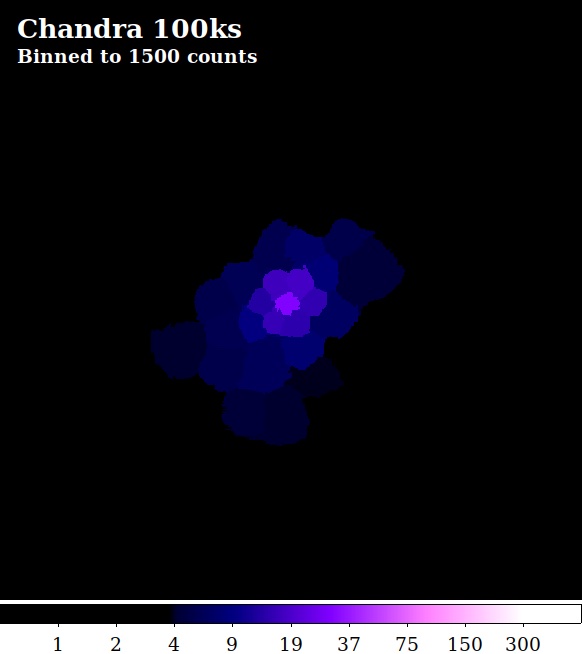}
\includegraphics[width=0.3\textwidth]{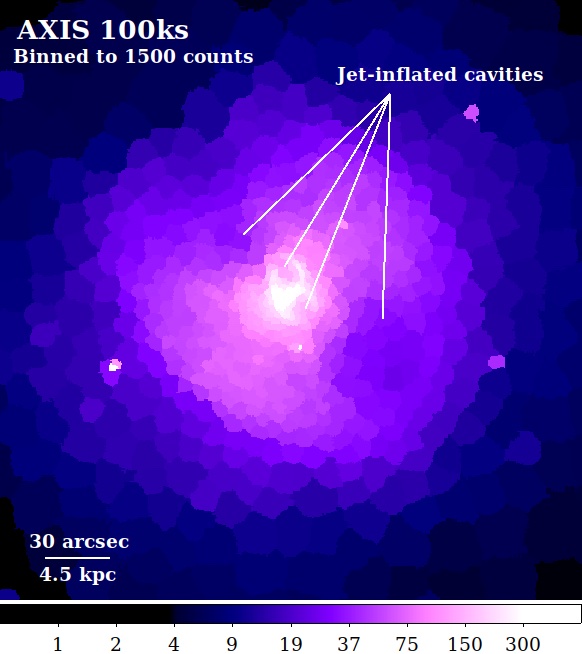}
\includegraphics[width=0.3\textwidth]{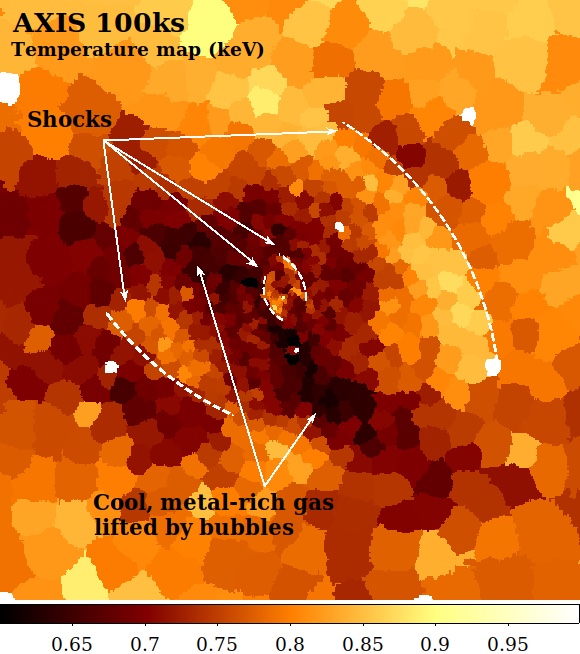}
\end{tabular}
\end{center}
\caption{Upper panels: AXIS will determine the stellar feedback energy input to galactic ecosystems by measuring their X-ray luminosity function in nearby galaxies. AXIS 0.3-2 keV, 150 ks image of M83 (left) with zoom region of an HII region/star cluster (center).  This survey will determine the X-ray luminosity function in more than 5000 clusters (right) down to or below the “knee” seen in H$\alpha$. Lower panels: Images of AGN feedback in NGC 5813.  The radio jet has carved out large cavities in the galaxy’s hot atmosphere, which are visible as depressions in X-ray surface brightness.  Left and center: Chandra (at launch) and AXIS simulated images of NGC 5813 with spatial binning for 1500 counts per region.  Color bars are matched and units are counts arcsec$^{-2}$.  Right: Simulated temperature map in keV for the AXIS observation, which probes the thermodynamic properties on the scales of the bubble rims, weak shocks, cold fronts, and cool gas filaments.  With 1500 counts per region, AXIS will map the gas properties to an accuracy of e.g. a few per cent in temperature.}\label{fig:feedback}
\end{figure} 

Stellar feedback can operate through several channels, including hot gas, cosmic rays, radiation, and mechanical feedback.  However, we lack an accurate energy budget that delineates these channels as a function of the mass, age, and environment of the star-forming region, leaving models for stellar feedback poorly constrained. AXIS imaging spectroscopy of nearby galaxies (Figure~\ref{fig:feedback} upper panels) will permit the characterization (luminosity, temperature, pressure, and metallicity) of thousands of massive star clusters, allowing us to trace the internal energy remaining in the hot gas, X-ray binaries, and cosmic-ray electrons.  The results will transform our picture of stellar feedback.

AGN also have a profound impact on galaxies.  Although the central SMBH has a mass of only $\sim 0.1$\% of its host galaxy, the enormous energy released as this relativistic object grows can slow or even completely suspend the growth of the entire galaxy \cite{fabian:12}. Prescriptions for AGN feedback are included in all cosmological simulations but need to be tuned in an ad hoc manner to obtain the observed galaxy population and its evolution--- neither the basic theory of AGN feedback nor the observational constraints on the feedback process are yet able to define these prescriptions from first principles.  In these models, it is AGN feedback that is primarily responsible for transforming MW and more massive gas-rich star-forming galaxies into red-and-dead galaxies, and maintaining that state.

AXIS will quantify how energetic AGN outflows, both accretion disk winds and relativistic jets, impact the surrounding interstellar/circumgalactic medium (ISM/CGM) in a representative sample of nearby galaxies.  In rapidly accreting systems (i.e., luminous AGN), the intense radiation field and high-velocity winds from the central accretion disk can shock-heat, photoionize, and expel gas from the surrounding ISM \cite{silk:98,fabian:99,king:03,murray:05}.  Models predict that the critical outflow/ISM interaction zone has scales of 100-1000\,pc, corresponding to 1-10\,arcseconds at a distance of 20\,Mpc.  Lower-luminosity AGN can also have a major impact on their host galaxy via their relativistic jets.  Jets dump energy directly into the hot (X-ray emitting) ISM/CGM of the host galaxy, inflating kpc-scale cavities of relativistic plasma and driving both shocks and turbulence in the hot medium.  For both low- and high-luminosity AGN feedback, the low X-ray surface brightness of this interaction zone means that CXO (the only current mission with the spatial resolution to disambiguate this emission from the bright AGN core) can only obtain a crude spectral characterization of the processes at play in the nearest systems \cite{fabbiano:18,maksym:19,trindade-falcao:23}. With the combination of high-spatial resolution, throughput, low background, and improved spectral resolution, AXIS will allow high-quality imaging spectroscopy of this region in nearby luminous AGN, allowing the complex physics of the AGN-ISM interaction to be disentangled.

A significant uncertainty plaguing our models of AGN feedback is the dependence with cosmic time; when in the history of the Universe did feedback processes start to strongly affect the properties of galaxies and galaxy clusters?  An important consequence of feedback is the enrichment of the intergalactic medium (IGM) by iron and other metals. This, in turn, leads to the observed universal enrichment of the intracluster medium (ICM) of galaxy clusters.  Exploiting exquisite sensitivity, low background, and the ability to disambiguate the ICM emission from embedded point-like AGN emission, AXIS will determine the epoch of ICM enrichment by tracking the evolution of ICM metallicity in the most massive galaxy clusters from the current time ($z<0.5$) through cosmic noon ($z=0.5-2.5$) and into cosmic morning ($z\sim 3$).  In addition to metallicity, AXIS imaging spectroscopy will determine the profile of thermodynamic qualities (temperature, pressure, and entropy) in the ICM as well as the presence of AGN-blown cavities, thereby providing direct probes of both instantaneous and integrated intracluster AGN feedback.

\subsection{WORLDS AND SUNS IN CONTEXT: Stellar Activity in Planet Formation}

{\it How do different stars affect the formation and evolution of their exoplanets? How do habitable environments arise and evolve within the context of their planetary systems?} 

Why do we live around a G-type star? This simple and fundamental question has perplexed astronomers for over a century.  Humans are children of our yellow Sun, yet lower-mass red stars (M or K stars) are ten times more numerous.  So what are the chances that intelligent life would develop around our comparatively rare star?  

\begin{figure}[!ht]
\begin{center}
\begin{tabular}{c} 
\hspace{-0.2cm}\includegraphics[trim=5 0 5 10,clip,width=\textwidth]{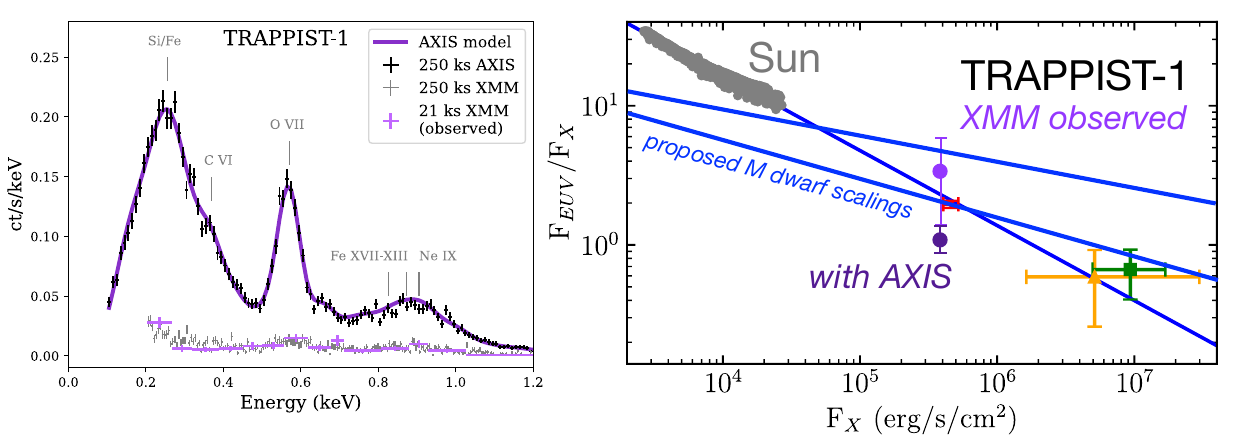}\\
\includegraphics[trim=10 5 5 10,clip,width=0.5\textwidth]{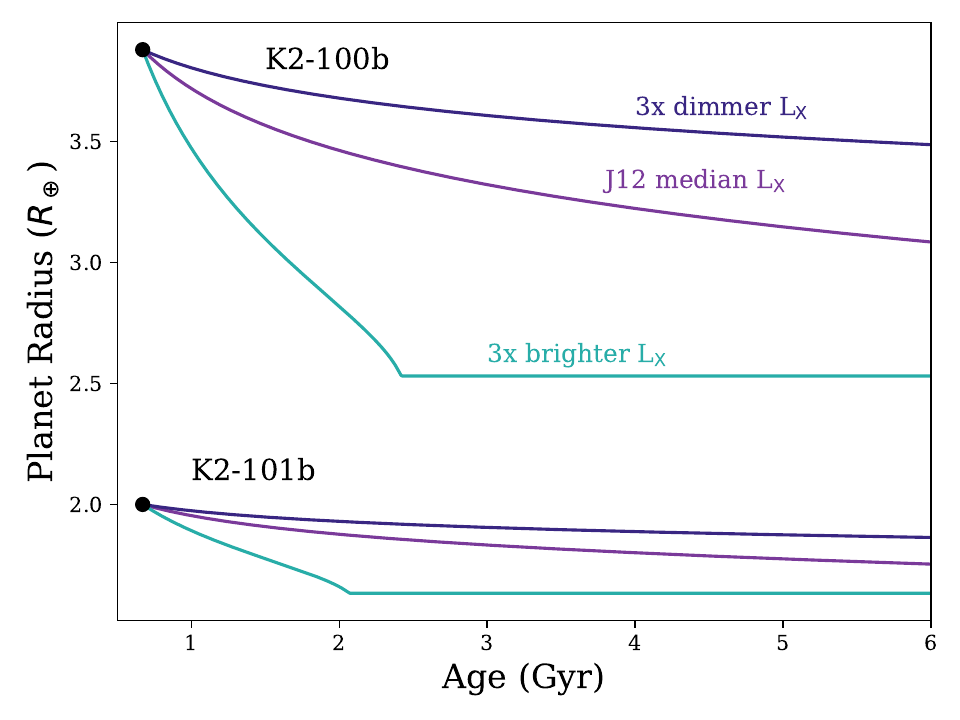}
\includegraphics[trim=10 5 5 10,clip,width=0.5\textwidth]{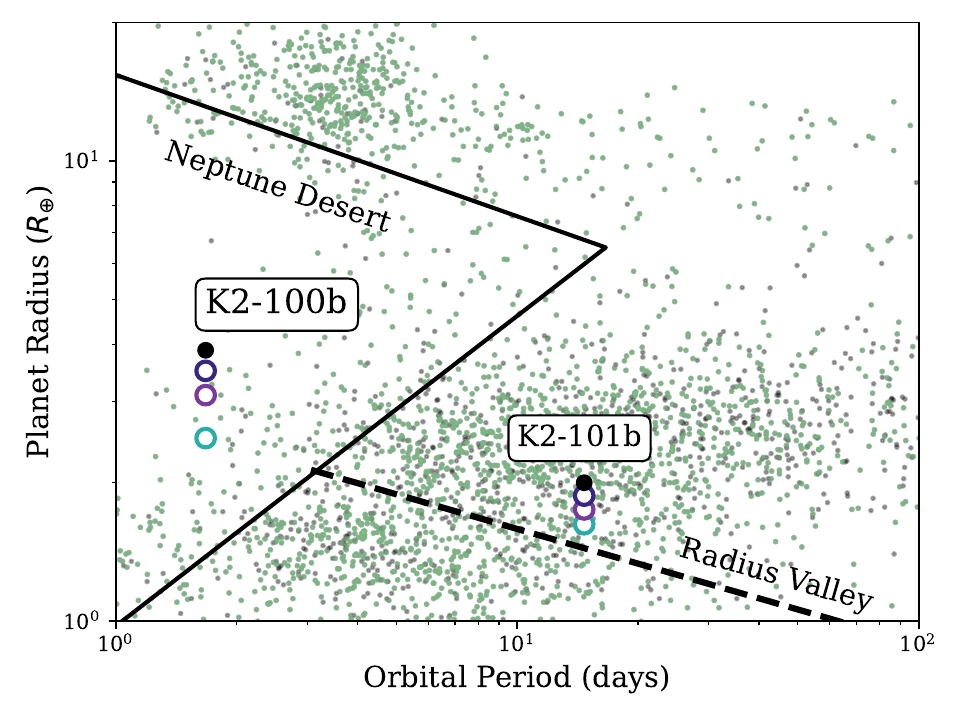}
\end{tabular}
\end{center}
\caption{{\textit{Top Left : }}Simulated 250~ks spectra for the M~dwarf TRAPPIST-1 \cite{Wheatley:2017}. {\textit{AXIS will provide a more exquisite X-ray signal and spectral features than a comparable observation with XMM}}, providing strong constraints on the differential emission measure (DEM) reconstruction required to fully understand the stellar atmosphere and activity level. A 250~ks monitoring campaign, executed over the course of AXIS's nominal mission lifetime, would also capture the flare activity and coronal mass ejections that could dramatically affect the atmospheric properties and habitability of the terrestrial planets orbiting these M dwarfs.
{{Top Right,} modified from King et al.~2018\cite{king:18} :} The EUV to X-ray scaling relation inferred from Solar data is overlaid with other proposed stellar scaling relations \cite{Johnstone:2021,SanzForcada:2011,sanz-forcada:22} and the measured XUV properties of three M dwarfs (red, yellow, and green)\cite{Chadney:2015}. The purple data points show the inferred TRAPPIST-1 EUV/X-ray luminosity ratio obtained via DEM reconstruction with a single X-ray luminosity data point versus a 250~ks observation with AXIS (dark purple). {{The soft X-ray sensitivity and energy resolution of AXIS improve accuracy and reduce EUV flux uncertainty from a factor of two to just 30\%}}, constraining EUV to X-ray scaling relationships and profoundly affecting our ability to predict surface habitability of terrestrial worlds.
{{Bottom Left:}} The predicted radius evolution under photoevaporative escape models for two representative Neptune/sub-Neptune planets in the young (670~Myr) Praesepe star cluster \cite{king:22}.  
The radius of both planets is expected to evolve significantly from their current radii (circle markers), due to photoevaporative escape. Different evolutionary tracks are shown, depending on where the X-ray luminosity of the host stars fall within the known dispersion (factor of $\sim$10) of empirical relations \cite{Jackson:2012}. {{AXIS observations of M-dwarf planet hosting stars will pin down their X-ray luminosities and variability}}, permitting more confident predictions of the impacts of stellar activity on planet atmospheres over stellar lifetimes.
{{Bottom Right:}} The observed radius versus orbital period of transiting planetary systems.
Solid black circles indicate the current radius and orbital period of K2-100b and K2-101b. The open circles predict the radius evolution to a system age of 6~Gyr, depending on the X-ray luminosity of their host stars (same color scheme as figure at left). {{AXIS observations will help us understand how planets evolve within this essential parameter space.}}
} 
\label{fig:exoplanets}
\end{figure} 

A leading hypothesis is that the coronal/flaring activity of the host star can play a critical role in habitability.  Cool stars such as M dwarfs have fully convective atmospheres, leading to magnetic/flaring activity that is more common than that observed in the stellar corona of the Sun. This magnetic/flaring activity generates X-ray and extreme-UV (XUV) emission \cite{Maggio:2023} that impacts the atmospheres of orbiting planets\cite{Vida:2017,Johnstone:2019}.  Harsh stellar environments have the ability to strip the entire planet of its atmosphere \cite{Lammer:2003,OwenJackson:2012}, or could dramatically alter the atmospheric and surface chemistry \cite{Lammer:2007}, prohibiting (or perhaps even catalyzing!) prebiotic chemical pathways \cite{GomezDeCastro:2021}.  However, to date the effect of stellar host activity on the structure and evolution of the planetary atmospheres remains one of the biggest uncertainties in our understanding of exoplanet atmosphere formation and habitability. 

With the launch of JWST, we have opened a new window to study exoplanet atmospheres. These studies are rapidly becoming a major field in their own right, but the complementary work needed to understand the effect of the stellar host on the planets and their atmospheres remains woefully behind. Although there have been important attempts to estimate the total XUV emission and flare contributions, the results are inconclusive, even conflicting: for instance, the few simultaneous X-ray and UV observations performed show that some X-ray and UV flares are not correlated\cite{Osten:2005}, and empirically establishing the relationship between flares observed in XUV and coronal mass ejections remains elusive for stars other than the Sun \cite{Aarnio:2011}. Therefore, we cannot robustly measure the total XUV energy nor the impact of coronal mass ejections imparted by the exoplanet host from UV observations alone. AXIS will transform this picture.  

The soft X-ray sensitivity and spectral resolution of AXIS will provide high signal-to-noise spectra of the coronal activity in key planetary systems known to host terrestrial planets in the habitable zone. Deep (250\,ks) exposures on M dwarfs known to host such planets will enable a full differential emission measure (DEM) solution, reconstructing the coronal plasma distribution with temperature, and hence providing the most reliable prediction for the EUV environment of the terrestrial planet, even when the EUV emission itself is completely extincted by Galactic dust (Figure~\ref{fig:exoplanets} top panels).  

AXIS will broaden these findings to the overall Galactic population by characterizing the evolution of stellar activity with mass and age in stellar clusters, ranging in age from Myr to Gyr.  To date, X-ray studies have found an order of magnitude spread in X-ray luminosity for stars of a given stellar type and age \cite{Jackson:2012, Wright:2018, Johnstone:2021}.  Through high-sensitivity cadenced observations of star clusters during the AXIS prime mission, we will determine whether this spread is a consequence of variability or whether otherwise identical stars can be consistently ``X-ray active'' or ``X-ray inactive''.  Due to its spatial resolution and sensitivity over a wide field of view, AXIS will measure the X-ray luminosities in a wide range of stellar types and ages and, very importantly, will also quantify their dispersion, through cadenced monitoring of stellar clusters.  By better understanding the evolution of X-ray luminosity, its variability, and contributions from stellar flares and coronal mass ejections for stars over their lifetimes, we can reconstruct the mass loss history for a given planet (Figure~\ref{fig:exoplanets} bottom panels). 

\subsection{NEW MESSENGERS AND NEW PHYSICS: The Explosive Ends of Stars}\label{subsec:tdamm}

{\it What are the progenitors of core-collapse supernovae? What are the progenitors of Type Ia supernovae? How are relativistic jets launched after the merger of two neutron stars?} 

Explosive transients associated with the end of a star's life, supernovae and neutron star mergers, are fundamental to a wide range of areas in astrophysics; they generate heavy elements and disperse them throughout galaxies; they inject energy and momentum into galaxies, regulating star formation and evolution; and Type Ia supernovae are important standardizable candles for measuring the accelerating expansion of the universe \cite{riess:98,perlmutter:99}, a discovery that led to the 2011 Nobel Prize in Physics.  Yet, despite their importance, we fundamentally have yet to firmly identify the progenitor properties of $>$50\% of supernovae and do not know what (or whether a) compact object forms after many of these explosive events.  AXIS will be a powerful facility for exploring these explosive facilities, achieving the Astro2020 high-priority recommendation for a “space-based time-domain and multi-messenger capability.”  AXIS provides fast ($<2$\,hr) and high-sensitivity follow-up to any sufficiently interesting transient source that has been localized to $<1\,{\rm degree}^2$. Concurrently, the AXIS transient alert module (TAM) will operate continuously during the mission and allow real-time discovery of X-ray transients, with source localizations and basic characteristics (flux, rise time, spectral hardness) disseminated to the community in $<10$\,minutes. Finally, the ability to conduct fast surveys of the Galactic Plane offers an unprecedented opportunity to discover the progenitors of explosive transients (particularly pre-merger binary white dwarfs and magnetars).

The AXIS-TAM will determine the nature of the progenitors of 100 core-collapse supernovae through real-time discoveries of supernova shock breakouts -- the moment when the first electromagnetic radiation escapes from the exploding star -- and will alert the greater astronomical community within 10 minutes.  Currently, only a single well-established example of a real-time X-ray shock breakout detection is known: SN2008D, a Type Ib (He-poor) supernova in the nearby (d = 27 Mpc) NGC2770. Although upcoming wide-field X-ray observatories such as the Einstein Probe will likely uncover more such nearby examples, the unprecedented sensitivity of AXIS will enable the detection of SN2008D-like events out to z $\sim 2.5$, near the peak of cosmic star formation \cite{madau:14}. As a result, AXIS will measure the evolution of progenitor properties, as well as the stripped-envelope core-collapse supernova rate, in an entirely unique manner. These discoveries are entirely serendipitous and do not require dedicated observing time.  The prompt multi-wavelength follow-up enabled by the rapid AXIS alerts will characterize the subsequent shock cooling emission, as well as the radioactively powered phase.

AXIS will determine the origin of Type Ia supernovae by increasing the sample of accreting ultracompact binaries (the most numerous sources of gravitational waves in the LISA era) by a factor of $>30$ and measuring their orbital evolution and merger rate. While it is widely agreed that Type Ia supernovae, our cosmic distance ladders, arise from the explosion of a white dwarf, the mechanism that causes the white dwarf to explode has been debated for decades. There are two main proposed channels: the so-called single-degenerate scenario, where a massive white dwarf near the Chandrasekhar mass accretes from a non-degenerate companion, and the double-degenerate scenario that results from the merger of two white dwarfs. To understand whether double degenerates are responsible for a large fraction of Type Ia supernovae, we need to [1] find more short-period, mass-transferring white dwarf (double-degenerate) binaries near the point of merger, which can only be done reliably in X-rays, and [2] determine whether those systems will indeed merge due to gravitational wave emission to become Type 1a supernovae or other types of white dwarf merger products, or whether mass transfer will stabilize their orbits, salvaging these systems from an explosive demise. 

\begin{figure} [t]
\begin{center}
\begin{tabular}{c} 
\includegraphics[width=0.9\textwidth]{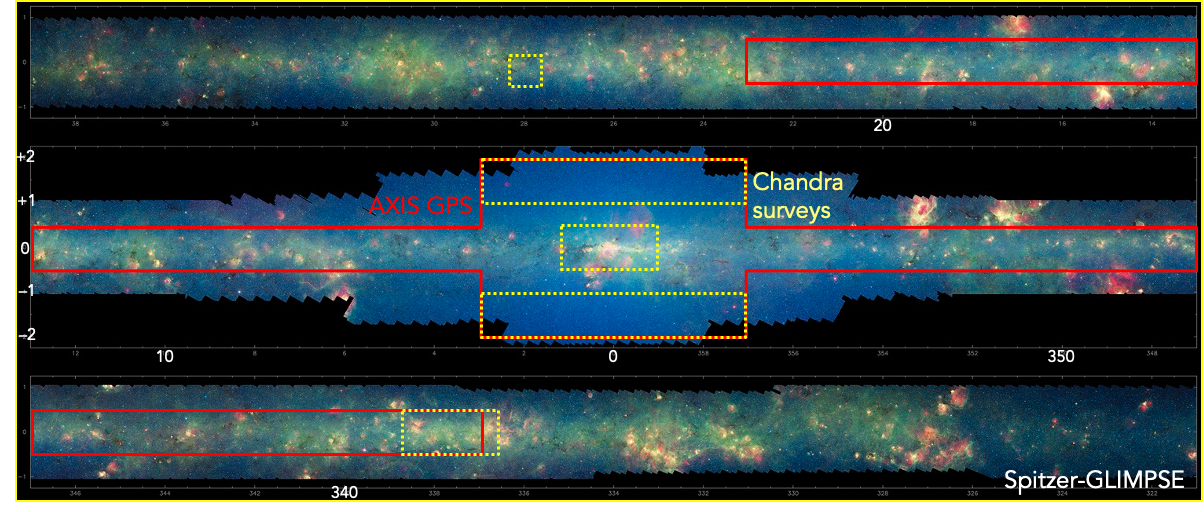}
\end{tabular}
\end{center}
\caption{The footprint of the AXIS Galactic Plane Survey, in red, overlaid on a {\it Spitzer} infrared backdrop. The {\it Chandra} survey fields are shown in yellow.  Observing regions probe an order of magnitude deeper than Chandra surveys with sub-arcsec X-ray positions for counterpart matching and provide X-ray coverage of the inner Milky Way’s obscured, star-forming regions revealed in the Spitzer-GLIMPSE survey.
}\label{fig:gps}
\end{figure} 

AXIS will conduct a sensitive time domain survey of the Galactic Plane to reveal this hidden population of ultra-short period ($<$10 minutes) mass-transferring double white dwarfs, increasing the known population from two to at least 60. AXIS will discover systems similar to HM Cancri, a direct impact accreting system that was discovered with ROSAT, due to its proximity (2 kpc \cite{munday:23}) and because its favorable inclination leads to a 100\% modulation of its X-ray flux. We estimate that there should be approximately $2 \times 10^{3}$ mass transferring and inspiraling double white dwarfs in the Galaxy. With our Galactic plane survey, we expect to be sensitive to essentially all such systems out to 8\,kpc, allowing us to probe approximately 10\% of such systems in the Galaxy. We estimate that of the $\sim$200 systems in our GPS footprint, about a third will be eclipsing and will exhibit strong X-ray periodicity among the million sources detected in the GPS.

AXIS will uncover the geometry of more than 50 new neutron star binary mergers detected via gravitational waves, including unique constraints on the angular structure of their relativistic jets, the system inclination, the merger environment, and elusive kilonova afterglows. Colliding neutron stars emit bursts of electromagnetic and gravitational radiation with the exemplar being the LIGO/Virgo source GW170817\cite{abbott:17a}.  Each messenger provides unique insights into the physics of the merger and its ability to drive relativistic outflows. Gravitational wave detections yield distances, progenitor masses, and spins, whereas the evolution of any relativistic jets detected via X-ray and other multi-wavelength observations probe the system inclination and the conditions of the surrounding interstellar medium. X-ray observations also powerfully discern the structure of the relativistic jet and the merger remnant, and offer a closer look at kilonova afterglows, outflows that trace the energetics of the central explosion and constrain the synthesis of heavy elements. By triggering off localized GW sources, AXIS will obtain light curves and spectra for more than 50 binary neutron star mergers with peak X-ray fluxes $\sim 3 \times 10^{-16}$ erg cm$^{-2}$ s$^{-1}$ (per 10 ks exposure), creating the first X-ray population study of these unique sources.

The Astro2020 Decadal Survey identified gravitational-wave multi-messenger astronomy as one of the top science drivers for astrophysics, stating that the New Messengers and New Physics theme “will exploit the new observational tools of gravitational waves and particles, along with temporal monitoring of the sky across the electromagnetic spectrum and wide-area surveys\...” Indeed, Astro2020 recommends that “in space, the highest-priority sustaining activity is a space-based time-domain and multi-messenger program of small- and medium-scale missions.” AXIS expertly fills this role.

\section{OBSERVATORY ARCHITECTURE}

\subsection{Overview of Architecture}
\label{subsec:arch_overview}

\begin{figure} [h]
\begin{center}
\begin{tabular}{c} 
\includegraphics[width=1\textwidth]{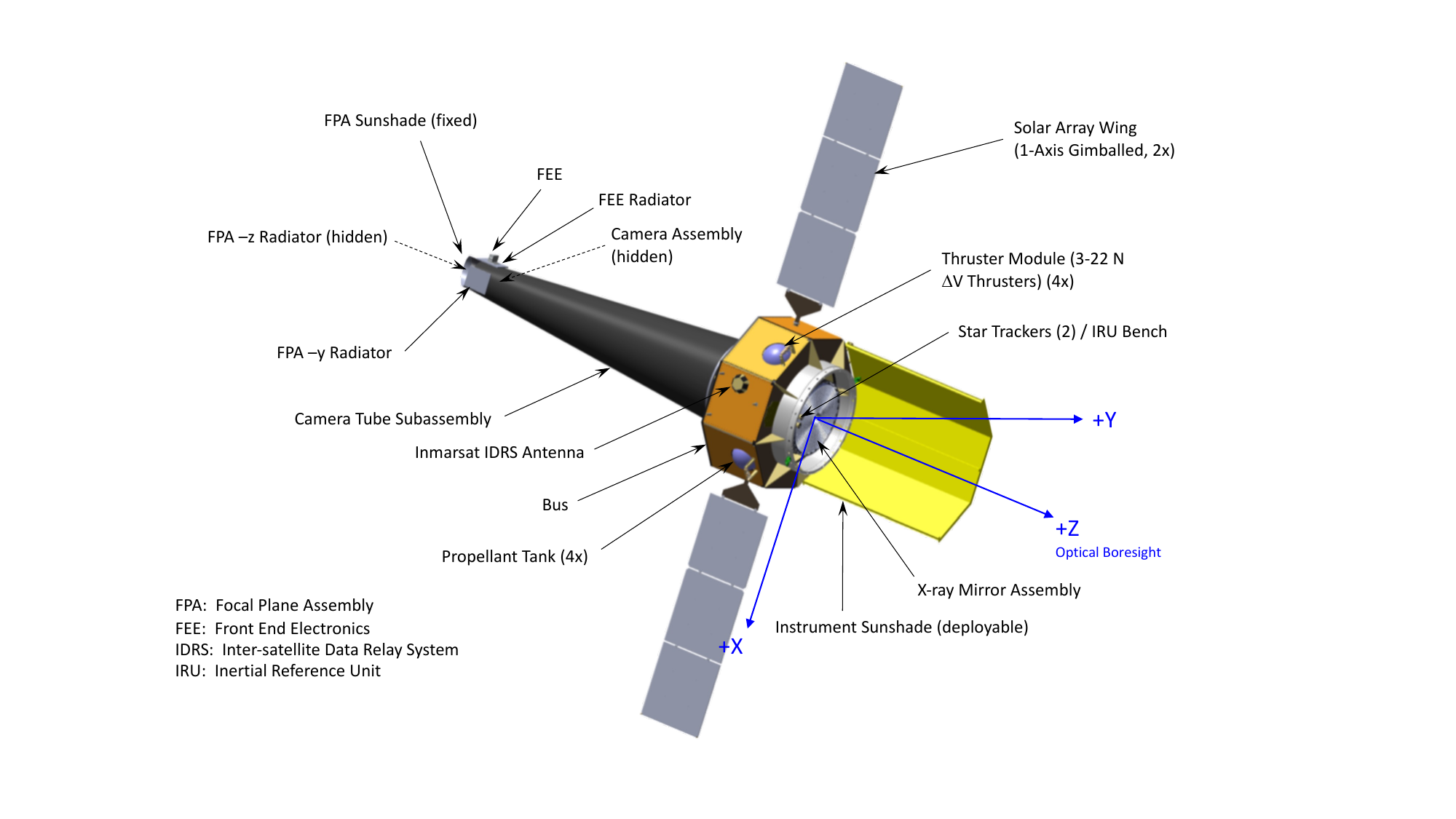}
\end{tabular}
\end{center}
\caption{The AXIS observatory in its deployed configuration. The total length from the edge of the aperture sunshade (yellow) to the back of the FPA sunshade is 13.8\,m, and the total span of the solar arrays is 16\,m. The outer surfaces of the instrument tube assemblies (black) will be covered by thermal blankets as well as micro-meteorite and orbital-debris protection (not shown).}\label{fig:observatory}
\end{figure} 

The guiding philosophy of the AXIS Team has been to design a powerful observatory optimized for the deep and fast imaging and spectroscopy described in Section~\ref{sec:science} while minimizing system complexity and implementation risk. The result is a single-instrument design with a single optical assembly (XMA) and a single detector array (FPA).   The XMA and FPA are mounted into a two-piece optical bench (the aperture tube and camera tube sub-assemblies, respectively) connected by a titanium splice ring which provides the interface to the spacecraft bus.  A deployable sunshade protects the XMA from both the optical and thermal load of direct sunlight.  At the other end of the instrument, the radiators associated with the cooling of the CCD arrays are protected from sunlight by a fixed sunshade, plus appropriate angling.  

AXIS utilizes a high-heritage, single-fault-tolerant spacecraft bus provided by Northrop Grumman.  A deployable one-axis gimballed solar array provides the AXIS instrument with adequate power without constraining the pointing directions. The spacecraft also provides communications.  Telescope slewing and pointing control is achieved via reaction wheels in the spacecraft, with occasional momentum dumping via magnetic torquers.  The spacecraft propulsion system is used only to correct any orbit insertion errors to achieve the initial 670\,km altitude orbit, enable end-of-mission controlled de-orbiting, and for collision avoidance maneuvers.

Mechanical and thermal stability, as well as accurate and precise pointing control and knowledge, are crucial for enabling the reconstruction of high-resolution imaging and spectroscopy that underpins all AXIS science. These considerations are woven through the AXIS design.  First, the XMA must be thermally stable to maintain image quality.  We employ active thermal control at the module level to maintain a constant temperature of $22\pm 0.25^\circ$C.  The modules are mounted into a holding structure (the ``spider'') constructed from a low coefficient of thermal expansion (CTE) composite and mechanically isolated from the spacecraft to the maximum extent possible. As a result, mechanical and thermal distortions of the XMA contribute $<0.25$\,arcsec (in quadrature) to the final PSF. Second, the telescope must maintain focus as AXIS orbits from the day-to-night side of the Earth and slews across the sky. The telescope tube sub-assemblies have exterior thermal blanketing and are also constructed from low-CTE composite. As a result, the CCD plane can remain within the $20$\,microns depth of focus of the XMA (which has a 9m focal length) without the need for active thermal control.  The whole FPA assembly is mounted on a Tip-Tilt-Focus-Mechanism (TTFM), which will achieve the initial focus during post-launch commissioning and can be used occasionally to remove secular trends in the XMA-FPA distance. Since the TTFM carries the weight of the FPA, a launch lock subassembly protects it from damaging loads during launch.

AXIS is a photon counting instrument that performs high-fidelity image reconstruction after-the-fact in ground processing by registering and stacking photons.  Thus, we require knowledge (not control), with $\sim 0.2$\,arcsec accuracy, of the sky coordinates of each detected photon. This breaks down into several steps, which again have driven the design of AXIS.  Star-trackers and an inertial reference unit (IRU) attached directly to the XMA provide knowledge of the instantaneous boresight direction, unaffected by any flexing between the spacecraft bus and AXIS instrument. A metrology system monitors the motions between the XMA and FPA due to tube distortions on the same cadence as the CCD readout. Finally, low-noise reaction wheels and fast CCD readout ($>5$ frames per second) ensure that spacecraft jitter does not irreducibly blur the reconstructed frames.

\subsection{X-ray Mirror Assembly (XMA)}\label{subsec:xma}

\begin{figure} [h]
\begin{center}
\begin{tabular}{c} 
\includegraphics[width=0.9\textwidth]{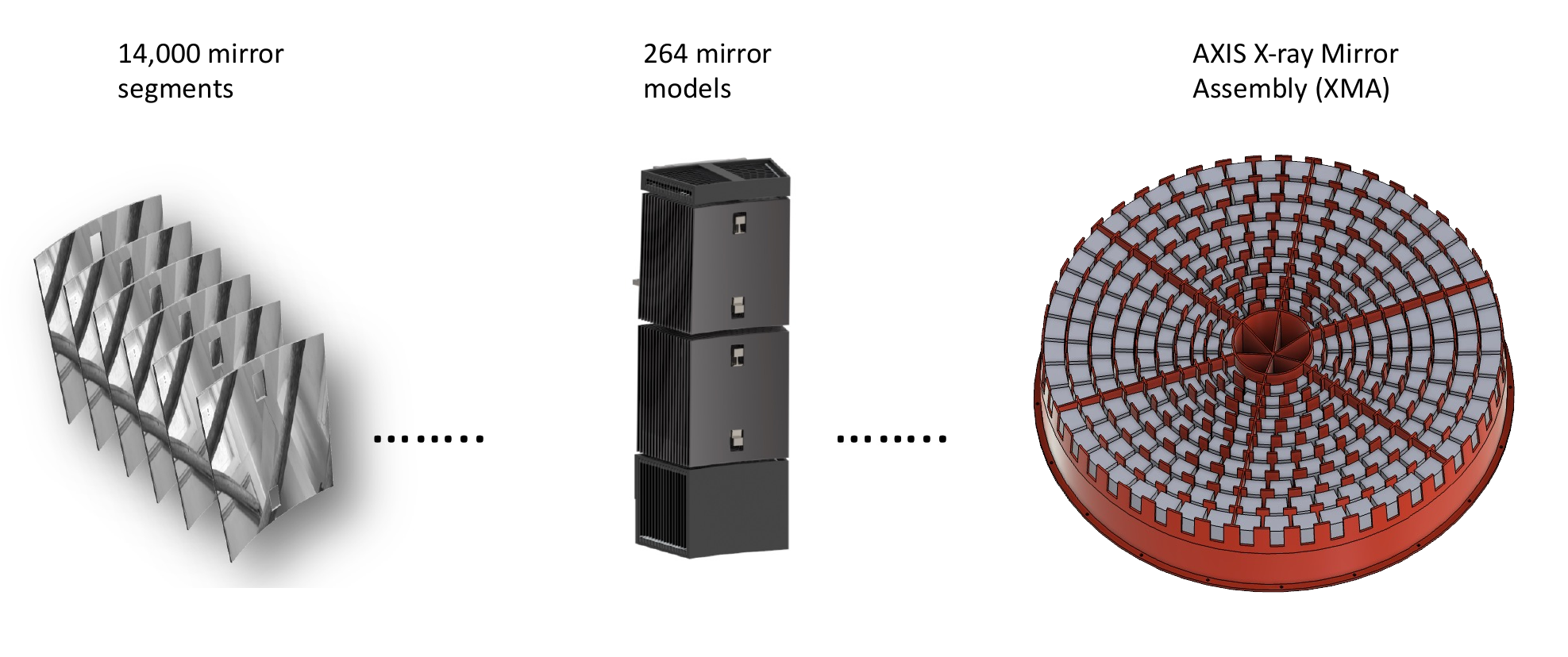}
\end{tabular}
\end{center}
\caption{The three major steps of manufacturing the XMA. Left: production and qualification of approximately 14,000 mirror segment; Middle: Alignment and integration of these mirror segments into 264 mirror modules, each of which is tested for both science performance and spaceflight environment; and Right: the mirror modules are assembled into the final AXIS XMA shown here integrated into the low-CTE composite ``spider'' structure with a 1.8\m outer diameter.}\label{fig:xma}
\end{figure}

The production of the XMA uses a new technology that has been under development by the Next Generation X-ray Optics (NGXO) team NASA's Goddard Space Flight Center. This technology, which incorporates knowledge and lessons learned from past X-ray telescopes such as Chandra, XMM-Newton, Suzaku, and NuSTAR, is based on a combination of precision polishing, alignment, and mass production to simultaneously achieve image quality, light weight, and low cost. A key innovation has been the use of single-crystal silicon as the basic substrate, which minimizes internal stresses and subsequent distortions of each mirror segment.  Figure~\ref{fig:xma} illustrates the three major steps that lead to a completed XMA.

In the first step, silicon mirror segments are mass-produced and certified by optical metrology for both their optical properties and structural integrity. The production of the mirror segments starts with a block of single-crystal silicon commercially procured, and ends with an atomic layer deposition step that coats the mirror segment, on both the front and back surfaces, with a thin iridium film that enhances their X-ray reflectivity and preserves their optical figure. Between these steps are generating, lapping, slicing, polishing, trimming, slotting, acid etching, and optical measurements. The entire mirror segment production process uses both unique technologies specifically developed for the purposes at hand and commercially-off-the-shelf technologies and equipment. These mirror segments are $>10$ times lighter and $>10$ times less expensive  but  of comparable optical quality, to Chandra's mirror shells. These combined characteristics serve as the foundation for manufacturing the AXIS XMA to meet the three-fold requirement of PSF precision, light weight, and production cost and schedule.

In the second step, the $\sim 14,000$ mirror segments are assembled into 264 mirror modules, each of which, for all intents and purposes, is an X-ray telescope in its own right. The assembly process consists of two major technical elements. The first technical element is an alignment process in which each mirror segment is properly manipulated to optimize its location and orientation relative to its conjugate. The details of the process include the use of temporary structures that hold the mirror segment without distortion that allow it to be manipulated with a state-of-the-art hexapod under the guidance of an optical beam that reads out the mirror segment's position and orientation. Following the successful alignment of the mirror segment while it is temporarily held, the second technical element is to permanently bond the mirror segment to the module structure with epoxy and two flexures. After the epoxy cures, the temporary structures are de-bonded from the mirror segments and removed, leaving in place the mirror segment permanently attached to the module. In general, the cure process results in changes in mirror alignment because epoxy tends to shrink in unpredictable ways. These changes are compensated for by adjusting two actuators that have been designed as part of the flexures. Once the adjustments are completed and their results verified by optical measurements, the actuators are staked permanently, preventing further movement.

In the final step, the mirror modules are aligned and bonded onto a superstructure that will structurally support the modules and serve as the structural interface between the mirror assembly and the spacecraft bus. This can be accomplished using standard engineering practices due to the relatively loose optical tolerance of aligning and permanently fixing the mirror modules. In general, the optical tolerance of co-aligning mirror modules is 10 to 100 times looser than that of aligning mirror segments. As such, the buildup of the XMA from mirror modules is largely a structural engineering exercise similar to those of co-aligning the three mirror modules of the XMM-Newton telescopes. Substantially similar work has been done many times in the past for other missions. 

The realization of the AXIS XMA requires the interplay and optimization of three major developments: technology, engineering, and production. The technology development is the maturation of all the technical elements required to make mirror modules. Engineering development consists, among other things, in incorporating all technical elements to design and build modules that meet all requirements of the spaceflight environment. Production development is to incorporate both technology and engineering development for implementation of an efficient mass manufacturing process leading to an XMA that not only meets all science performance and spaceflight environment requirements but also project schedule and cost requirements. 

The single-crystal silicon X-ray mirror technology has demonstrated its potential for making an XMA that meets, and likely exceeds, AXIS requirements. For example, many mirror segments have been fabricated that are better in figure quality than those of the Chandra mirrors. Furthermore, a single pair of mirror segments has already been aligned, bonded, and X-ray tested, achieving an image quality of
0.8\,arcsec HPD, comparable to CXO mirror's performance and close to AXIS requirements. Recently, mirror modules with two pairs of mirror segments have been built and X-ray tested, achieving an image quality of 2.1\,arcsec HPD, passing qualification-level vibration tests. In the coming months, the NGXO team expects to combine all of these elements into mirror modules that will simultaneously meet image quality and spaceflight environment requirements. Meanwhile, a study is underway to canvass industry expertise and equipment to design and implement a mass production process that will meet all AXIS requirements, including schedule and cost.

\subsection{Focal Plane Assembly}\label{subsec:fpa}

This AXIS Focal Plane Assembly (FPA)\cite{Milleretal2023} is designed to take advantage of the high throughput and spatial resolution of the XMA, although these very capabilities present challenges. The AXIS camera must operate much faster than similar instruments on heritage missions (CXO and Suzaku) while retaining excellent spectral imaging performance. The FPA incorporates an array of fast-readout charge-coupled devices (CCDs) designed and fabricated by MIT Lincoln Laboratory (MIT/LL) and building on a long line of successful space instruments spanning the last three decades. Each CCD is coupled with an application-specific integrated circuit (ASIC) specifically designed by Stanford University to provide low-noise, low-power amplification of the CCD analog signal. The front-end electronics incorporate modern digital processing to further reduce noise, and the back-end electronics under development at Penn State implement a state-of-the-art FPGA-based Event Recognition Processor (ERP) to greatly reduce the telemetry stream and a Transient Alert Module (TAM) to detect changes in flux among sources in the field of view and rapidly disseminate transient alerts to the community. 

\begin{figure}[h]
\begin{center}
\includegraphics[width=.4\linewidth]{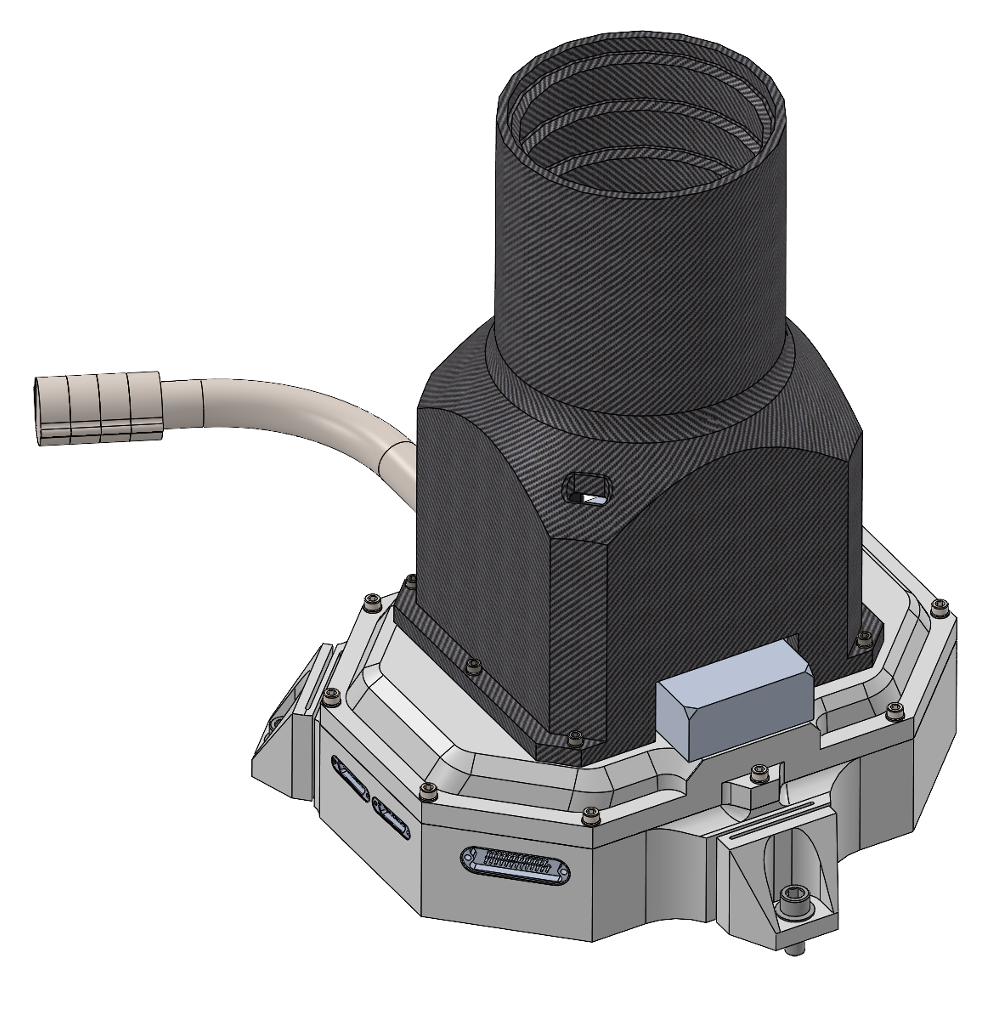}
\hspace*{.5in}
\includegraphics[width=.35\linewidth]{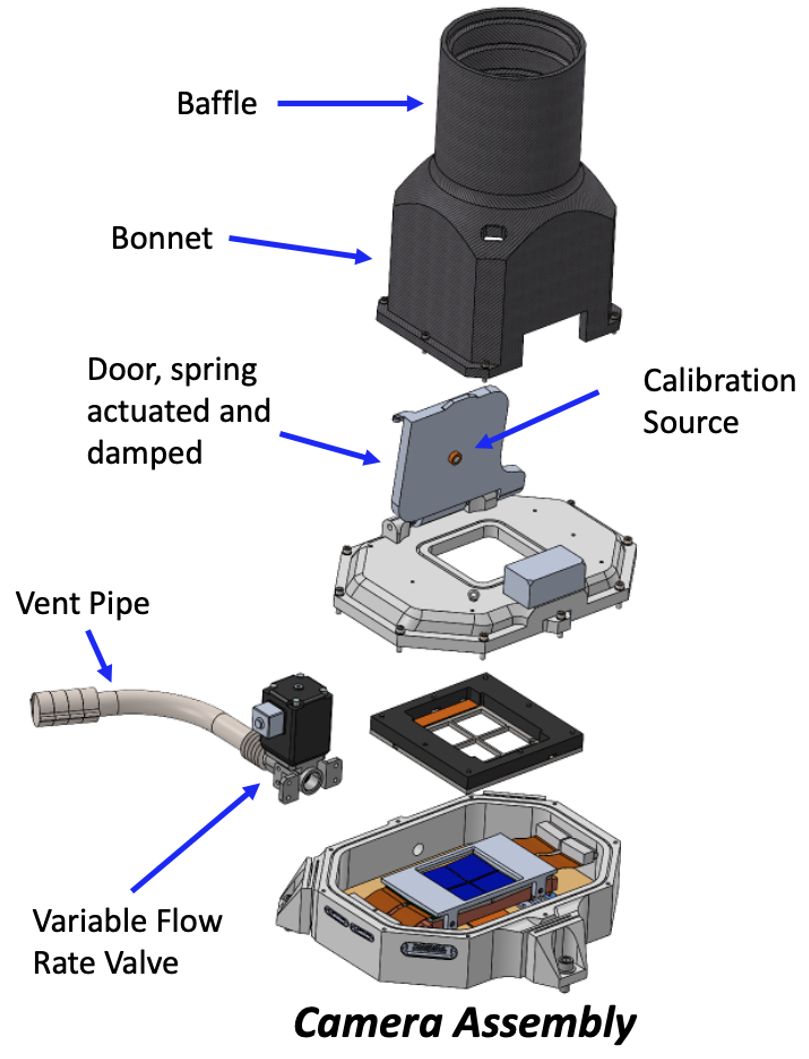}
\end{center}
\caption{The AXIS Focal Plane Assembly (FPA), rendered as integrated (left) and in an exploded view (right) with various features labeled. The CCD detectors can be seen as dark blue squares in the detector housing at the lower right.}
\label{fig:housing}
\end{figure} 

The baseline frame rate of 5 frames per second (fps) is sufficient to avoid pile-up on most sources in a typical pointing. For eight CCD outputs and ASIC channels per detector running at 2 MHz, and a parallel transfer speed of 1 MHz, the current best estimate for focal plane full-frame readout rate is 7 fps. Increasing the frame rate is possible through design changes, such as increasing the number of outputs; or through operational changes, such as increasing the output rate to 5 MHz. This latter change may come at the cost of increased noise and reduced soft X-ray response, and so could be implemented as a choice based on the science target. Similarly, pile-up of bright point-like sources could be reduced by reading out only a small sub-array of the aimpoint detector or even employing a continuous-clocking mode that eliminates all spatial information along CCD columns. Such a scheme would also improve time resolution to the sub-ms regime. We are confident that we can meet the AXIS goal of 20 fps with some combination of design and operational considerations, while retaining good spectral performance to allow ground-breaking science.

The detector system is housed in a reliable, high-heritage camera structure that builds on lessons learned from previous missions (\ref{fig:housing}). In particular, the housing is kept under vacuum during all stages of ground testing and integration, in part to prevent the accumulation of molecular contamination, which can migrate and adhere to the cold ($-90^{\circ}$C) CCD surfaces and degrade the soft X-ray sensitivity. The housing is vented to space once on orbit, and a warm ($+20^\circ$C) contamination blocking filter further prevents molecular contamination from building up along the optical path.

Key components of the FPA, including the CCD detectors and ASICs, have been undergoing technology development for several years and are on track to meet or exceed the AXIS mission requirements. A summary of current performance along with more details about the FPA camera design can be found elsewhere in these proceedings.\cite{Milleretal2023}

\section{MISSION IMPLEMENTATION}

AXIS will be launched from the Kennedy Space Center in Q2 CY2032 into a circular LEO with an inclination $<8$\,degree and an initial altitude of 670\,km (98-minute orbital period).  After the initial commissioning of the spacecraft and instrument, AXIS will conduct a 5-year prime science mission.  Analysis shows that the altitude remains above 610\,km for the full duration of this prime mission even using very conservative assumptions for the solar irradiation and corresponding atmospheric drag.  The desirable range of operational altitudes (610--670\,km) is bounded above by the worsening radiation environment and below by observing efficiency and ground-station considerations.

\subsection{Concept of Operations}

AXIS is fundamentally a pointed observatory.  The majority of the mission will be spent in ``Target mode'', executing 500--5000\,s observations of astrophysical targets/fields to a pre-planned schedule. Each orbit, AXIS will observe the target field until either (i) the field becomes occulted by the Earth, (ii) AXIS enters the SAA, or (iii) the necessary total exposure time has been collected.  Once one of these stopping criteria have been satisfied, AXIS will slew to the next target and begin a new observation.  The observing schedule will be optimized to maximize the observation time subject to timing/cadence constraints imposed by the scientific investigations.  SAA passages will be used to conduct large-angle slews that are sometimes necessary to set up some longer observations, while non-SAA slews will attempt to minimize the slew angle/time.  The reaction wheels on the spacecraft enable fast slewing, with conservative/worst-case estimates of $<9.5$\,mins for a 120\,degree slew (including settling time).  With this slew speed and the optimized observing plan, simulations of on-orbit operation show that we can achieve $>74\%$ observing efficiency, not including time set aside for calibration.

As described in Section~\ref{subsec:tdamm}, AXIS is also designed to be a powerful observatory for time-domain astronomy.  The AXIS Science Operations Center (AXIS-SOC) will interface with Transient Broker Networks (TBN) to receive real-time alerts from a variety of other facilities (including the Rubin Observatory, ground-based gravitational wave observatories, high-energy gamma-ray and cosmic ray observatories, and neutrino detectors). Alerts will be filtered by automated algorithms according to pre-defined criteria aimed at identifying the most scientifically interesting events that are sufficiently localized and accessible to AXIS (i.e. not in the Sun/anti-Sun exclusion zone). The filtering process will be tuned to accept $\sim 1$ alert per day.  Accepted events and the short-term revision to the observing plan will be passed to the AXIS Mission Operation Center (AXIS-MOC) and then uplinked to AXIS via a commercial L-band service.  Ultimately, in this ``Target of Opportunity" (ToO) Mode, AXIS will be observing the new target in $<2$\,hours from the receipt of the alert at the SOC and until Earth occultation or SAA passage.  Normal Target Mode resumes once observations of the ToO target end.  Science data for the ToO target is downlinked in the next regular ground-station contact, resulting in an ultimate latency of $<12$\,hours from receipt of the alert in the SOC to the availability of initial ``quick look'' science data.

The TAM operates at all times during science operations and will trigger on defined thresholds that identify new sources or sources that have varied dramatically from their baseline in the AXIS field.  When triggered, a data package is prepared containing the source location (accurate to $<2$\,arcsec), flux, rise-time and spectral hardness and downlinked via a low-latency commercial L-band service.  In this ``Transient Event Alert'' (TEA) mode, the basic information of the source can be communicated to the community via the AXIS-SOC and TBN in $<10$\,minutes from initial detection.   If additional pre-defined criteria are satisfied, the AXIS-SOC can itself respond to the AXIS-TEA by triggering an ``AXIS ToO Mode'' follow-up observation which would facilitate an immediate return by AXIS to the field in question.

\subsection{Programmatics}

AXIS is a response to NASA's Astrophysical Probe Explorer (APEX) program with major partners being the University of Maryland (Principal Investigator Institute), MIT (Deputy-PI and FPA-lead), Stanford and Penn State (FPA and Back-End Electronics), NASA Goddard Space Flight Center (Project Management and XMA-Lead), Northrop Grumman (spacecraft and metrology), and the University of Erlangen (science simulation software).  If selected, Phase B mission formulation begins in Q2-CY2026 and launch occurs in Q2-CY2032. The cost-cap is FY23\$1B, not including launch and costs associated with the General Observer (GO) program which are separately funded by NASA.

The AXIS Team is charged with formulating and executing a compelling 5-year science program that addresses Astro2020 priorities and justifies the investment in the mission; this is the program described in Section~\ref{sec:science}. The APEX peer review process and ultimate mission selection are based on the compelling nature of these proposed science objectives and the feasibility of successfully achieving these objectives.  However, at least 70\% of the science time in the prime mission will be open to the General Observer (GO) community through a competitive peer-reviewed proposal process managed by NASA.  With the enormous discovery space enabled by the capabilities of AXIS, we expect the GO community to both build upon and broaden the scientific investigations of the AXIS Team, and conduct fundamentally new strands of study, some of which have yet to be conceived.

At the conclusion of the 5-year prime mission, a review of spacecraft and instrument heal th will be conducted and inform any decision by NASA to approve/fund an extended mission phase.  There are no expendables or orbit considerations that preclude a 10-year science mission, although the legal requirement for controlled de-orbit of this large observatory may necessitate truncation of the extended mission if the spacecraft control systems begin to degrade.

\section{CONCLUSIONS}

The need for a large collecting area, high-spatial resolution X-ray observatory with moderate-resolution spectroscopic capability has long been recognized, but is now becoming a critical need if the full promise of JWST and Astro2020 science is to be realized. For the past two decades, the principal barrier to such an observatory has been the limitations of the X-ray optics and the lack of programmatic opportunities.  Building on more than two decades of strategic technology development at NASA-GSFC, and empowered by colossal investments by the semi-conductor industry, the single-crystal segmented X-ray optics developed by NGXO finally allow us to make this leap to light-weight, robust, high-resolution optics in a Probe mission.  The result is AXIS --- an elegant, single-instrument observatory with unprecedented sensitivity across a field-of-view that is almost the size of the full Moon, augmented with capabilities that make it a powerful time-domain facility.  

This paper has presented an  overview of both the implementation of the mission and the AXIS Team Science Program.  If selected, AXIS will launch in the first half of 2032 and conduct a 5-year prime mission that will tackle core questions raised by Astro2020.  History gives us a clear lesson however --- with the enormous discovery space opened by AXIS, the most important contributions of AXIS to physics and astronomy will be by answering questions that have yet to even be conceived.

\acknowledgments 

C.S.R. and R.F.M. gratefully acknowledge support of the AXIS Program from the Department of Astronomy, College of Mathematical and Natural Sciences, and Vice President for Research Office at the University of Maryland via the Big Opportunity Fund.

The MIT focal plane team gratefully acknowledges support from NASA through the Strategic Astrophysics Technology (SAT) program, grants 80NSSC18K0138, 80NSSC19K0401 and 80NSSC23K0211 to MIT, and from the Kavli Research Infrastructure Fund of the MIT Kavli Institute for Astrophysics and Space Research. Stanford team members acknowledge support from NASA through Astrophysics Research and Analysis (APRA) grants 80NSSC19K0499 and 80NSSC22K1921, and from the Kavli Institute for Particle Astrophysics and Cosmology.

D.H. acknowledges support from the Canada Research Chairs (CRC) program and the National Sciences and Engineering Research Council of Canada (NSERC) Discovery Grant program.

K.W.C acknowledges the support by NASA through the CRESST-II Cooperative Agreement under award number 80GSFC21M0002.

HRR acknowledges support from an STFC Ernest Rutherford Fellowship and an Anne McLaren Fellowship.

S.S.H. acknowledges support from NSERC through the Canada Research Chairs program and the Discovery Grants program, and from the Canadian Space Agency.

F.E.B. acknowledges support from ANID-Chile BASAL CATA FB210003, FONDECYT Regular 1200495,
and Millennium Science Initiative Program  – ICN12\_009

T.D. acknowledges support from the McDonnell Center for the Space Sciences at Washington University in St. Louis.

\bibliography{axisref} 
\bibliographystyle{spiebib} 

\end{document}